\documentclass[nofootinbib,showpacs,showkeys]{revtex4}
\usepackage{graphicx,color}
\usepackage{amssymb}
\usepackage{amsmath}
\usepackage{bm}
\usepackage{lipsum}
\usepackage{latexsym}
\usepackage{dcolumn}
\usepackage{float}
\usepackage{hyperref}

\begin{document}

\title{\boldmath Matter Growth in Imperfect Fluid Cosmology}

\author{Winfried Zimdahl}
 \email{winfried.zimdahl@pq.cnpq.br}
\affiliation{ PPGCosmo \& N\'{u}cleo Cosmo-UFES, Universidade Federal do Esp\'{\i}rito Santo,
Av. Fernando Ferrari, 514, Campus de Goiabeiras, CEP 29075-910,
Vit\'oria, Esp\'{\i}rito Santo, Brazil}
\author{Hermano Velten}%
 \email{hermano.velten@ufop.edu.br}
\affiliation{Departamento de Física, Universidade Federal de Ouro Preto (UFOP), CEP 35400-000, Ouro Preto, MG, Brazil}
\author{William C. Algoner}%
 \email{walgoner@fisica.ufpr.br}
\affiliation{Departamento de F\'{\i}sica, Universidade Federal do
Paran\'a, 81531-980 Curitiba, Brazil}


\begin{abstract}Extensions of Einstein's General Relativity (GR) can formally be given a GR structure in which additional geometric degrees of freedom are mapped on an effective energy-momentum tensor.
The corresponding effective cosmic medium can then be modeled
as an imperfect fluid within GR.
The imperfect fluid structure allows us to include, on a phenomenological basis, anisotropic stresses and energy fluxes which are considered as potential signatures for deviations from the cosmological standard
$\Lambda$-cold-dark-matter ($\Lambda$CDM) model.
As an example, we consider the dynamics of a scalar-tensor extension of the standard model, the $e_{\Phi}\Lambda$CDM model.
We constrain the magnitudes of anisotropic pressure and energy flux with the help of redshift-space distortion (RSD) data for the matter growth function $f \sigma_8$.

\keywords{fluid cosmology; cosmological perturbation theory; scalar-tensor theory; cosmic structure~formation}
\end{abstract}

\maketitle

\section{Introduction}
\label{introduction}

The last two decades have  seen a tremendous activity within the community of cosmologists and astrophysicists to find a satisfactory explanation for the results of the observations of supernovae of type Ia (SNIa),
reported in \cite{Acc1,Acc2,Acc3}, which suggested an accelerated expansion of the scale factor of the Robertson--Walker (RW) metric. The apparently simplest way to account for this behavior is to assume the existence of a cosmological constant $\Lambda$ with a suitable value (see, e.g., \cite{amendolaTsuji,ellisMaartensMcC}).
From a purely general relativistic (GR)
point of view, $\Lambda$ might be seen as another gravitational constant along with Newton's gravitational constant.
On the other hand, an effective cosmological constant has been associated with the quantum vacuum (see, e.g.,~\cite{jeromeM}). Taking this into account, the actually measured gravitational constant
might be a combination of a purely geometric and a quantum contribution.
This context gave rise to a number of discussions concerning the cosmic coincidence problem  \cite{DECCP1,DECCP2,DECCP3,DECCP4,coincidence}. 

Since accelerated expansion in inflationary models of the early universe is conveniently described in terms of scalar fields,
a corresponding  description has also been applied to the current phase of the cosmic evolution. This is equivalent to making the cosmological ``constant" a dynamic quantity.
Along this line, a number of different approaches have been developed which either assume the existence of some form of exotic matter within GR or modify Einstein's theory of gravity. In the meantime, a huge amount of data of various types have been accumulated \cite{planck18,des18,ska18,ligo}.

Facing the multitude of models at hand, one might wish to keep an eye on unifying phenomenological aspects of different types of description.
It is this intention which motivates the present paper.
Our starting point is the observation that any extension of Einstein's GR can formally be recast into an Einsteinian structure with a  suitably defined effective energy-momentum tensor \cite{madsen,pimentel}.  In the presence of an adequate timelike vector field, this energy-momentum tensor is then characterized by an energy density, a scalar isotropic pressure, an anisotropic pressure and an energy flux.
It has the structure of the energy-momentum tensor of an imperfect fluid \cite{battye,battye13,battye16,faraoni18}.
Imperfect-fluid dynamics can provide an effective description for cosmological models beyond the standard model.
Anisotropic stresses and/or energy fluxes are typical ingredients  of a non-standard dynamics of, e.g., scalar-tensor  theories.
A robust phenomenological theory may represent a unifying framework for approaches that differ in their detailed underlying microscopic dynamics.
Since the anisotropic pressure determines the gravitational slip, a suitable parametrization may be used to quantify potential deviations from the standard model and, combined with observational  data, to put limits on such deviations. A similar comment holds for heat--flux caused deviations.

The resulting gravitational dynamics depends on the (effective) fluid quantities directly, but it does not explicitly depend on the detailed underlying microscopic dynamics which is either unknown or beyond a straightforward and transparent analytical treatment \cite{LAmend}.
For a scalar field, e.g., there is a dependence on this field only through the mentioned fluid type quantities,
in the simplest case energy density and isotropic pressure, but there is no further direct dependence on the scalar-field details themselves.
On this basis, we aim to give a phenomenological fluid-type description of the cosmological dynamics in which different models are determined by a set of phenomenological parameters such as an equation-of-state parameter and  an adiabatic sound speed parameter. Additionally, and this is what we consider to be the new aspect of our study, it is necessary to introduce analogous parameters for the anisotropic pressure and the energy flux.
We split the total effective energy-momentum tensor  into two components, one of them being a pressureless perfect fluid, the second one an imperfect fluid. The pressureless fluid is supposed to account for some form of (dark) matter, the imperfect component is intended to provide an effective description of dark energy.
While rather general, such a purely phenomenological description necessarily leaves open the microscopic origin of these quantities.

One of the tools to discriminate between different models of the cosmological dark sector is the growth rate of matter perturbations \cite{nesseris08,basilakos12,huterer15,nesseris15,boss}.
Competing models with similar behavior in the homogeneous and isotropic background  will generally have different predictions for the dynamics of matter inhomogeneities.
As an example, we consider the simplest possible scalar-tensor extension of the $\Lambda$CDM model.
In this minimalist approach, we remain in the vicinity of the standard model at the present epoch \cite{WHW}.
The present paper advances a preliminary study in \cite{WHW2} where the matter growth rate was obtained in a simplified manner on the basis of a rough approximation which both avoided a solution of the full dynamics and neglected the heat flux.
Here, we consider the full perturbation dynamics with the heat flux included.

In Section \ref{EMT}, we introduce the energy-momentum tensors of the cosmic medium as a whole and those of the individual components.
The general conservation laws are presented in Section \ref{Conservation}.
Section~\ref{Perturbation} is devoted to the perturbation dynamics. The relevant metric and fluid quantities are introduced and a modified Poisson equation is obtained.
A coupled system of two second-order equations for the entire cosmological dynamics is established from which the dynamics of matter perturbations is derived.
All this is applied in Section \ref{ephilambda} to the special case of the $e_{\Phi}\Lambda$CDM model.
In Section \ref{data}, we constrain the parameters that describe deviations from the standard model by contrasting the predictions of this model with recent redshift-space distortion (RSD)
data.
Section \ref{discussion} provides our conclusions concerning the status of the present approach.

\section{Energy-Momentum Tensors}
\label{EMT}

We assume the cosmic medium to consist of pressureless perfect-fluid type matter (subindex ($m$)) and a second component, modeled as a general imperfect fluid (subindex ($x$)).
The cosmic substratum as a whole is then an imperfect fluid as well. It is described by an energy-momentum tensor
\begin{equation}\label{Ttot}
T_{\mu\nu} = T_{(m)\mu\nu} + T_{(x)\mu\nu}.
\end{equation}
Greek tensorial indices run over $0,\ 1,\ 2,\ 3$.
The first part, $T_{(m)\mu\nu}$, describes pressureless matter through
\begin{equation}\label{}
T_{(m)\mu\nu} = \rho_{m}u_{(m)\mu}u_{(m)\nu},
\end{equation}
where $u_{(m)}^{\mu}$ with $u_{(m)}^{\mu}u_{(m)\mu}=-1$ is the matter four-velocity which may be associated with an observer. The matter energy density is
\begin{equation}\label{}
\rho_{m} = T_{(m)\mu\nu}u^{\mu}_{(m)}u^{\nu}_{(m)}.
\end{equation}
(For scalar quantities, we use the subindex $m$ instead of ($m$)).
The second part, $T_{(x)\mu\nu}$, is given by the general split with respect to a timelike unit vector $u_{(x)}^{\mu}$,
\begin{equation}\label{}
T_{(x)\mu\nu} = \rho_{x}u_{(x)\mu}u_{(x)\nu} + p_{x}h_{(x)\mu\nu} + \Pi_{(x)\mu\nu}
+ q_{(x)\mu}u_{(x)\nu} + q_{(x)\nu}u_{(x)\mu},
\end{equation}
where $u_{(x)}^{\mu}u_{(x)\mu} = -1$ and
\begin{equation}\label{}
\rho_{x} = T_{(x)\mu\nu}u_{(x)}^{\mu}u_{(x)}^{\nu}, \quad p_{x} = \frac{1}{3}h_{(x)}^{\mu\nu}T_{(x)\mu\nu},\quad
q_{(x)\mu} = - h_{(x)\mu}^{\nu}T_{(x)\nu\sigma}u_{(x)}^{\sigma},
\end{equation}
\begin{equation}\label{}
\Pi_{(x)\mu\nu} = h_{(x)\langle\mu}^{\sigma}h_{(x)\nu\rangle}^{\tau}T_{(x)\sigma\tau}
\equiv \frac{1}{2}\left(h_{(x)\mu}^{\sigma}h_{(x)\nu}^{\tau}T_{(x)\sigma\tau}
+ h_{(x)\nu}^{\sigma}h_{(x)\mu}^{\tau}T_{(x)\sigma\tau}\right)
 - \frac{1}{3}h_{(x)\mu\nu} h_{(x)}^{\sigma\tau}
T_{(x)\sigma\tau}
\end{equation}
with
\begin{equation}\label{}
h_{(x)}^{\mu\nu} = g^{\mu\nu} + u_{(x)}^{\mu}u_{(x)}^{\nu},\quad
h_{(x)}^{\mu\nu}u_{(x)\nu} =
q_{(x)\mu}u_{(x)}^{\mu} = \Pi_{(x)\mu\nu}u_{(x)}^{\mu} = \Pi^{\mu}_{(x)\mu} =0.
\end{equation}
(For scalar quantities, we use the subindex $x$ instead of ($x$)).
Here, $\rho_{x}$ is the energy density of component $x$ and $p_{x}$ is its isotropic pressure.
The heat--flux vector is denoted by $q_{(x)\mu}$ and the anisotropic pressure by $\Pi_{(x)\mu\nu}$.
In general, the timelike unit vector $u_{(x)}^{\mu}$ does not coincide with $u_{(m)}^{\mu}$. In a scalar-field representation, e.g., it may be related to the timelike gradient of a scalar field.
We shall choose both four-velocities to coincide in a homogeneous and isotropic background, but they will differ, in general, on the perturbative level.

For the total energy-momentum tensor, we assume a decomposition with respect to still another timelike unit vector $u^{\mu}$,
\begin{equation}\label{}
T_{\mu\nu} = \rho u_{\mu}u_{\nu} + ph_{\mu\nu} + \Pi_{\mu\nu}
+ q_{\mu}u_{\nu} + q_{\nu}u_{\mu},
\end{equation}
where
\begin{equation}\label{projrho}
\rho = T_{\mu\nu}u_{}^{\mu}u_{}^{\nu}, \quad p  = \frac{1}{3}h^{\mu\nu}T_{\mu\nu},\quad
q_{\mu} = - h_{\mu}^{\nu}T_{\nu\sigma}u^{\sigma},
\end{equation}
\begin{equation}\label{projpi}
\Pi_{\mu\nu} = h_{\langle\mu}^{\sigma}h_{\nu\rangle}^{\tau}T_{\sigma\tau}
\equiv \frac{1}{2}\left(h_{\mu}^{\sigma}h_{\nu}^{\tau}T_{\sigma\tau} + h_{\nu}^{\sigma}h_{\mu}^{\tau}T_{\sigma\tau}\right)
 - \frac{1}{3}h_{\mu\nu} h^{\sigma\tau}
T_{\sigma\tau},
\end{equation}
with
\begin{equation}\label{}
u^{\mu}u_{\mu} = -1,\quad h^{\mu\nu} = g^{\mu\nu} + u^{\mu}u^{\nu},\quad
h^{\mu\nu}u_{\nu} =
q_{\mu}u^{\mu} = \Pi_{\mu\nu}u^{\mu} = \Pi^{\mu}_{\mu} =0.
\end{equation}
In this decomposition, $\rho$ is the total energy density, $p$ is the total isotropic pressure, $q_{\mu}$ is the total heat flux and
$\Pi_{\mu\nu}$ is the total anisotropic pressure.
The relation between $u^{\mu}$, $u^{\mu}_{(m)}$ and $u^{\mu}_{(x)}$ will be clarified~below.

\section{General Conservation Equations}
\label{Conservation}

We shall assume separate energy-momentum conservation of both components.
The separate matter conservation equations are
\begin{equation}
-u_{(m)\nu}T^{\nu\kappa}_{(m);\kappa} = \rho_{m,\alpha}u_{(m)}^{\alpha} +  \Theta_{m} \rho_{m} = 0\
\label{eb1}
\end{equation}
and
\begin{equation}
h_{(m)\nu}^{\alpha}T^{\nu\kappa}_{(m);\kappa} = \rho_{m}\dot{u}_{(m)}^{\alpha} = 0.
\label{mb1}
\end{equation}
Here, $\Theta_{m} \equiv u^{\alpha}_{(m);\alpha}$ and $\dot{u}_{(m)}^{\alpha} \equiv u_{(m) ;\beta}^{\alpha}u_{(m)}^{\beta}$ is the matter four-acceleration.
In the homogeneous and isotropic background, Equation~(\ref{eb1}) reduces to
\begin{equation}\label{mattercons}
\frac{d \rho_{m}}{dt} + 3H \rho_{m} =0,
\end{equation}
where $H$ is the Hubble rate $H\equiv \frac{\dot{a}}{a}$ and $a$ is the scale factor of the RW metric.
In this background, Equation~(\ref{mb1}) is identically satisfied.

The conservation equations for the x component are
\begin{equation}
-u_{(x)\nu}T^{\nu\kappa}_{(x);\kappa} = \rho_{x,\alpha}u_{(x)}^{\alpha} +  \Theta_{x} \left(\rho_{x} + p_{x}\right) + \nabla_{\mu}q_{(x)}^{\mu}= 0,
\label{eb2}
\end{equation}
where $\nabla_{\mu}q_{(x)}^{\mu} \equiv h_{\mu}^{\alpha}q_{(x);\alpha}^{\mu}$ and
\begin{equation}
h_{(x)\nu}^{\alpha}T^{\nu\kappa}_{(x);\kappa} = \left(\rho_{x} + p_{x}\right)\dot{u}_{(x)}^{a} + p_{x,\nu}h_{(x)}^{\alpha\nu}
+ \nabla^{\kappa}\Pi_{(x)\mu\kappa}
  + h^{\nu}_{\mu(x)}\dot{q}_{(x)\nu} +\frac{4}{3}\Theta h_{\mu\kappa(x)}q_{(x)}^{\kappa} = 0
\label{mb2}
\end{equation}
with $\Theta_{x} \equiv u^{\alpha}_{(x);\alpha}$ and a four-acceleration $\dot{u}_{(x)}^{\alpha} \equiv u_{(x) ;\beta}^{\alpha}u_{(x)}^{\beta}$.
We have neglected here terms that will not contribute in a linear perturbation theory about a homogeneous and isotropic background with vanishing $\Pi_{(x)\mu\kappa}$ and $q^{\mu}_{(x)}$.
The pressurefree matter motion is geodesic ($\dot{u}_{(m)}^{\alpha} = 0$), while the motion of the x component is generally not.
For the background dynamics, one has
\begin{equation}\label{}
\frac{d \rho_{x}}{d t} + 3H\left(1 + w_{x}\right)\rho_{x} = 0\
\end{equation}
with an equation-of-state (EoS) parameter $w_{x}$ that has to be specified for a given model.

The general total energy-conservation equation becomes (neglecting contributions of $\dot{u}_{\mu}q^{\mu}$
with $\dot{u}_{\alpha} \equiv u_{\alpha;\beta}u^{\beta}$
and $\sigma_{\mu\nu}\Pi^{\mu\nu}$ with $\sigma_{\mu\nu} \equiv \frac{1}{2}\left(u_{\mu;\nu}+ u_{\nu;\mu}-\frac{2}{3}h_{\mu\nu}u^{\alpha}_{\alpha}\right)$
which will be of higher than first order in a linear perturbation theory)
\begin{equation}
 \dot{\rho} +   \Theta\left(\rho + p\right) + \nabla_{\mu}q^{\mu} = 0,
\label{ebtot}
\end{equation}
where $\dot{\rho} \equiv \rho_{,\mu}u^{\mu}$ and $\Theta \equiv u^{\alpha}_{;\alpha}$.  For the momentum conservation, we have
(again taking into account only those terms that contribute in a linear perturbation theory)
\begin{equation}
\label{dotu}
  \left(\rho +p\right)\dot{u}_{\mu} + \nabla_{\mu}p + \nabla^{\kappa}\Pi_{\mu\kappa}
  + h^{\nu}_{\mu}\dot{q}_{\nu} + \frac{4}{3}\Theta h_{\mu\kappa}q^{\kappa}= 0.
\end{equation}
In the homogeneous and isotropic background, $q^{\mu} =\dot{u}_{\mu} = \Pi_{\mu\kappa} =\sigma_{\mu\kappa}= \omega_{\mu\kappa} \equiv \frac{1}{2}\left(u_{\mu;\kappa}- u_{\kappa;\mu}\right)=0 $.
In our linear perturbation theory, we will neglect products of these
first-order quantities.

In the homogeneous and isotropic background, we identify all four-velocities, i.e. $u_{(m)}^{a} = u_{(x)}^{a} = u^{a}$.
Then,
\begin{equation}\label{}
\frac{d \rho}{d t} + 3H\left(1 + w\right)\rho = 0\
\end{equation}
with
\begin{equation}\label{}
w \equiv \frac{p}{\rho} = \frac{p_{x}}{\rho} = w_{x}\Omega_{x},\quad
\Omega_{x} = \frac{\rho_{x}}{\rho}= 1 - \Omega_{m},\quad \Omega_{m} = \frac{\rho_{m}}{\rho}.
\end{equation}

\section{Perturbation Dynamics}
\label{Perturbation}
\subsection{Metric and Fluid Quantities}
We restrict ourselves to scalar perturbations, described by the line element (in the longitudinal gauge)
\begin{equation}
\mbox{d}s^{2} = - \left(1 + 2 \phi\right)\mbox{d}t^2 +
a^2\left(1-2\psi\right)\delta _{ab} \mbox{d}x^a\mbox{d}x^b .\label{ds}
\end{equation}
The fluid-dynamical system of perturbation equations will be established by adequately generalizing the corresponding steps in  \cite{Wiliam1,baVDF,ricci,alonso}.
First-order fluid quantities will be denoted by a hat symbol.
The perturbed time components  (tensorial index $0$) of the four-velocities are
\begin{equation}
\hat{u}_{0} = \hat{u}^{0} = \hat{u}_{(m)}^{0} =\hat{u}_{(x)}^{0}  = \frac{1}{2}\hat{g}_{00} = - \phi\ .
\label{u0}
\end{equation}
We define the (three-) scalar quantities $v$, $v_{m}$ and $v_{x}$ by (latin indices denote spatial components)
\begin{equation}
a^2\hat{u}^{m} = \hat{u}_{m} \equiv v_{,m},
\label{}
\end{equation}
as well as by
\begin{equation}
a^2\hat{u}^{a}_{(m)}  = \hat{u}_{(m) a} \equiv v_{m,a} \quad \mathrm{and}\quad
a^2\hat{u}^{a}_{(x)}  = \hat{u}_{(x)a} \equiv v_{x,a},
\label{vmvx}
\end{equation}
respectively.
The perturbed expansion scalar becomes
\begin{equation}
\Theta = u^{\mu}_{;\mu}\quad \Rightarrow\quad\hat{\Theta} = \frac{1}{a^2}\Delta v  -
3\dot{\psi} - 3 H\phi ,\label{Thetaexp}
\end{equation}
where $\Delta$ denotes the three-dimensional Laplacian.
The corresponding first-order expressions for $\Theta_{m} = u^{\mu}_{(m);\mu}$ and $\Theta_{x} = u^{\mu}_{(x);\mu}$ are
\begin{equation}
\hat{\Theta}_{m} = \frac{1}{a^2}\Delta v_{m}  -
3\dot{\psi} - 3 H\phi \quad \mathrm{and}\quad \hat{\Theta}_{x} = \frac{1}{a^2}\Delta v_{x}  -
3\dot{\psi} - 3 H\phi,
\label{Thetaexpl}
\end{equation}
respectively.

With $\hat{T}_{0}^{0} = \hat{\rho}$ as well as $\hat{T}_{(x)0}^{0} = \hat{\rho}_{x}$ and $\hat{T}_{(m)0}^{0} = \hat{\rho}_{m},$ it follows that, at first order,
\begin{equation}\label{}
\hat{\rho} = \hat{\rho}_{x} + \hat{\rho}_{m}.
\end{equation}
With
\begin{equation}\label{}
\hat{T}_{a}^{0} = \left(\rho + p\right)\hat{u}_{a}+q_{a}, \quad \hat{T}_{(m)a}^{0} = \rho_{m}\hat{u}_{(m)a}, \quad
\hat{T}_{(x)a}^{0} = \left(\rho_{x} + p_{x}\right)\hat{u}_{(x)a}+q_{(x)a},
\end{equation}
the first-order relation
\begin{equation}\label{}
\left(\rho + p\right)\hat{u}_{a}+q_{a}= \rho_{m}\hat{u}_{(m)a} + \left(\rho_{x} + p_{x}\right)\hat{u}_{(x)a}+q_{(x)a}
\end{equation}
is valid. According to our restriction to scalar perturbations, the vectors $q_{(x)a}$ and $q_{a}$ are represented by gradients of scalars, i.e.,
\begin{equation}\label{}
q_{(x)a} = q_{x,a}\quad \mathrm{and} \quad q_{a} = q_{,a},
\end{equation}
respectively. This leads to
\begin{equation}\label{sumvq}
v = \frac{\rho_{m}}{\rho +p}v_{m} +  \frac{\rho_{x}+p_{x}}{\rho +p}v_{x} +  \frac{q_{x}-q}{\rho +p},
\end{equation}
i.e., in the presence of heat fluxes the relation between the different velocities is modified compared with the perfect-fluid case.

From the perturbed spatial components
\begin{equation}\label{}
\hat{T}_{a}^{b}  = \hat{p}\delta_{a}^{b} + \Pi_{a}^{b},\quad \hat{T}_{(m)a}^{b} = 0,\quad \hat{T}_{(x)a}^{b}  = \hat{p}_{x}\delta_{a}^{b} + \Pi_{(x)a}^{b},
\end{equation}
we have
\begin{equation}\label{}
\hat{p}\delta_{a}^{b} + \Pi_{a}^{b}  =\hat{p}_{x}\delta_{a}^{b} + \Pi_{(x)a}^{b}.
\end{equation}
Since $\Pi_{a}^{a} = \Pi_{(x)a}^{a} =0,$ this reduces to
\begin{equation}\label{}
\hat{p} = \hat{p}_{x},\qquad \Pi_{a}^{b}  = \Pi_{(x)a}^{b}.
\end{equation}

\subsection{Modified Poisson Equation}

Taking into account heat--flux effects in the perturbation dynamics modifies the usual Poisson-type equation that relates the comoving density perturbations and the gravitational potential $\psi$.
Coupling  the 0-0 and the  0-a field equations leads to
\begin{equation}\label{}
\Delta\psi  = 4 \pi Ga^{2}\left(\hat{\rho}^{c} -3Hq\right)
= \frac{3}{2}H^{2}a^{2}\left(\frac{\hat{\rho}^{c}}{\rho} - 3\frac{Hq}{\rho}\right),
\end{equation}
where we introduced the comoving density perturbation
\begin{equation}\label{}
\hat{\rho}^{c} \equiv \hat{\rho} -3H\left(\rho +p\right)v.
\end{equation}
Then, the (generalized) Poisson equation becomes (in the $k$ space)
\begin{equation}\label{poissoneps}
 -\frac{2}{3}\frac{k^{2}}{H^{2}a^{2}}\psi = \varepsilon,
\end{equation}
where
\begin{equation}\label{epsdelta}
\varepsilon \equiv \delta - 3H\left[\left(1 + \frac{p}{\rho}\right)v +\frac{q}{\rho}\right].
\end{equation}
 The heat--flux contribution has to be included in the generalized definition of comoving energy--density perturbations.
 The quantity $\varepsilon$ represents a generalized comoving energy--density perturbation which takes into account also the existence of a heat flux contribution $q$ in addition to the velocity potential $v$.
Equation (\ref{poissoneps}) with (\ref{epsdelta}) generalizes the well-known corresponding relation for perfect fluids.
 Either the potential $\psi$ or the (generalized) comoving energy--density perturbation $\varepsilon$ may be used as independent variable of the perturbation theory.

\subsection{Combination of Conservation and Raychaudhuri Equations}
The basic set of first-order perturbation equations can be obtained by adequately generalizing previous fluid cosmological calculations  (cf. \cite{Wiliam1,baVDF,ricci,alonso}).
The essential ingredient is a combination of the total first-order conservation equations with the first-order Raychaudhuri equation. The result is
\begin{eqnarray}\label{epsprpr}
&&\varepsilon^{\prime\prime} + \left(\frac{3}{2} - \frac{15}{2}\frac{p}{\rho} + 3\frac{p^{\prime}}{\rho^{\prime}}\right)\frac{\varepsilon^{\prime}}{a}
 - \left[\frac{3}{2} + 12\frac{p}{\rho} - \frac{9}{2}\frac{p^{2}}{\rho^{2}} - 9\frac{p^{\prime}}{\rho^{\prime}}\right]\frac{\varepsilon}{a^{2}}
 - \frac{1}{a^2 H^{2}}\frac{\Delta}{a^{2}}\left(\frac{\hat{p}^{c}}{\rho} - \frac{p^{\prime}}{\rho^{\prime}}l\right)
 \nonumber\\
&&\qquad\qquad- \frac{3}{2}\left(1+\frac{p}{\rho}\right)\left[\frac{l^{\prime}}{a} + \frac{3}{2a^{2}}\left(1-\frac{p}{\rho}\right)l\right]\nonumber\\
&&\qquad\qquad\qquad - \frac{2}{a}\frac{\Delta\Pi^{\prime}}{a^{2}\rho}
- \left[3\left(1-\frac{p}{\rho}+2\frac{p^{\prime}}{\rho^{\prime}}\right)\frac{\Delta}{a^{2}H^{2}}
    +\frac{2}{3} \frac{\Delta^{2}}{a^{4}H^{4}}
    \right]\frac{H^{2}}{a^{2}}\frac{\Pi}{\rho}
=0,
\end{eqnarray}
where $l \equiv  \frac{\Theta q}{\rho}$ encodes the heat--flux contribution and the prime denotes a derivative with respect to the scale factor $a$.
So far, neither the pressure perturbations $\hat{p}^{c}$ nor the quantities $l$ and $\Pi$ are specified.
In a strict sense, they are to be determined from an underlying microscopic theory along with the EoS. If such theory is not available, progress can be made through a phenomenological approach.

\subsection{Relative Perturbations}

With the help of the quantities
\begin{equation}\label{}
D^{c} \equiv \frac{\hat{\rho}^{c}}{\rho + p} = \frac{\delta^{c}}{1+p/\rho},\quad \hat{\rho}_{m}^{c} \equiv \hat{\rho}_{m} + \dot{\rho}_{m}v,\quad
\delta^{c}_{m} \equiv \frac{\hat{\rho}_{m}^{c}}{\rho_{m}},
\end{equation}
we define the relative density perturbations (``entropy perturbations")
\begin{equation}\label{defSm}
S_{m} \equiv D^{c} - \delta^{c}_{m},
\end{equation}
as the difference between total and pure matter perturbations.
The perturbations of the pressure are the sum of an adiabatic part $\frac{\dot{p}}{\dot{\rho}}\hat{\rho}^{c}$ and a nonadiabatic contribution $\hat{p}_{nad}$,
\begin{equation}
\hat{p}^{c} = \frac{\dot{p}}{\dot{\rho}}\hat{\rho}^{c} + \hat{p}_{nad}.
\label{pc}
\end{equation}
For our configuration, the nonadiabatic part explicitly becomes
\begin{equation}
\hat{p}_{nad}
 =  \hat{p}_{x}^{c}-
\frac{\dot{p}_x}{\dot{\rho}_x}\hat{\rho}_{x}^{c}
+ \frac{\rho_{m} \left(\rho_{x} + p_{x}\right)}{\left(\rho +
p\right)} \frac{\dot{p}_x}{\dot{\rho}_x}
\left[\frac{\hat{\rho}_{x}^{c}}{\rho_{x} + p_{x}} -
\frac{\hat{\rho}_{m}^{c}}{\rho_{m}} \right],
\label{pna}
\end{equation}
where
\begin{equation}\label{}
\hat{\rho}_{x}^{c} \equiv \hat{\rho}_{x} + \dot{\rho}_{x}v \quad \mathrm{and} \quad
\hat{p}_{x}^{c} \equiv \hat{p}_{x} + \dot{p}_{x}v.
\end{equation}
The combination $\hat{p}_{x}^{c}-
\frac{\dot{p}_x}{\dot{\rho}_x}\hat{\rho}_{x}^{c}$ in Equation (\ref{pna}) accounts for the intrinsic nonadiabatic perturbations of the $x$ component while the last term appears due to the two-component nature of the cosmic medium. As a consequence, the fluid as a whole is nonadiabatic even if each of its components is adiabatic on its own.
With $\hat{\rho}_{x}^{c} = \hat{\rho}^{c} - \hat{\rho}_{m}^{c}$ in the last term on the right-hand side of
Equation (\ref{pna}), we may write
\begin{equation}\label{}
\frac{\hat{\rho}_{x}^{c}}{\rho_{x} + p_{x}} - \frac{\hat{\rho}_{m}^{c}}{\rho_{m}}
= \frac{\rho + p}{\rho_{x} + p_{x}} \left(D^{c} - \delta^{c}_{m} \right) = \frac{\rho + p}{\rho_{x} + p_{x}}S_{m}.
\end{equation}
Then, the nonadiabatic pressure perturbations are
\begin{equation}
\hat{p}_{nad}
 =  \hat{p}_{x}^{c}-
\frac{\dot{p}_x}{\dot{\rho}_x}\hat{\rho}_{x}^{c}
+ \rho_{m}\frac{\dot{p}_x}{\dot{\rho}_x}
S_{m}.
\label{pna2}
\end{equation}
This implies that, through the  pressure perturbations, the dynamics of the  total energy--density perturbation $\varepsilon$ (or $\delta^{c}$) is coupled to the dynamics of $S_{m}$
(cf. Equation (\ref{epsprpr})).
To obtain the dynamics of $S_{m},$ we combine the conservation equations for the total medium with those for the matter component.
On this basis, the following second-order equation for $S_{m}$ can be derived (again generalizing steps described in \cite{Wiliam1,baVDF,ricci,alonso}),
\begin{eqnarray}
S_{m}^{\prime\prime} +  \frac{3}{2}\left(1-\frac{p}{\rho}\right)\frac{S_{m}^{\prime}}{a} - \frac{\Delta}{a^{2}H^{2}}\frac{\hat{p}^{c}}{a^{2}\left(\rho +p\right)}
+ \frac{3}{a}\frac{\hat{p}^{\prime}_{nad}}{\rho +p} + 9\left(\frac{7}{6}+\frac{p^{\prime}}{\rho^{\prime}}  - \frac{1}{2}\frac{p}{\rho}\right)
\frac{\hat{p}_{nad}}{a^{2}\left(\rho +p\right)}&&\nonumber  \\
-\frac{2}{3}\frac{1}{a^{2}H^{2}}\frac{\Delta^{2}\Pi}{a^{4}\left(\rho +p\right)}
+ \frac{3}{a^{2}H^{2}}\frac{p^{\prime}}{\rho^{\prime}}\frac{H\Delta q}{a^{2}\left(\rho +p\right)}&=& 0.
\label{sprpr}
\end{eqnarray}
As in Equation~(\ref{epsprpr}), neither the pressure perturbations $\hat{p}^{c}$ nor the quantities $l$ (or $q$) and $\Pi$ are specified. The idea now is to determine the pressure perturbations as well as $l$ and $\Pi$ in terms of our basic variables $\varepsilon$ and $S_{m}$ in order to establish a closed system for $\varepsilon$ and $S_{m}$ from which subsequently the matter perturbations can be obtained.

\subsection{Perturbations of (an-)Isotropic Pressures and Energy Flux}

The pressure perturbations should be considered in the rest frame of the $x$
component. Since the combination
$\hat{p}_{x}-
\frac{\dot{p}_x}{\dot{\rho}_x}\hat{\rho}_{x}$ is gauge invariant, the first part on the right-hand side of Equation~(\ref{pna2}) may be written
\begin{equation}\label{}
\hat{p}_{x}^{c}-
\frac{\dot{p}_x}{\dot{\rho}_x}\hat{\rho}_{x}^{c} = \hat{p}_{x}-
\frac{\dot{p}_x}{\dot{\rho}_x}\hat{\rho}_{x} = \hat{p}_{x}^{c_{x}}-
\frac{\dot{p}_x}{\dot{\rho}_x}\hat{\rho}_{x}^{c_{x}},
\end{equation}
with
\begin{equation}\label{pxcx}
\hat{p}_{x}^{c_{x}} \equiv \hat{p}_{x} + \dot{p}_{x}v_{x}, \qquad \hat{\rho}_{x}^{c_{x}}\equiv \hat{\rho}_{x} + \dot{\rho}_{x}v_{x},
\end{equation}
where $v_{x}$ is the velocity potential of component $x$, introduced in Formula (\ref{vmvx}).

The generalized gauge-invariant comoving energy--density perturbation of the $x$ component is
\begin{equation}\label{}
\hat{\rho}_{x}^{\varepsilon} \equiv \hat{\rho}_{x} - 3H\left[\left(\rho_{x}+p_{x}\right)v_{x} + q_{x}\right].
\end{equation}
This definition for $\hat{\rho}_{x}^{\varepsilon}$ parallels the definition of the comoving energy--density perturbation $\varepsilon$ for the medium as a whole
(cf.Equation~(\ref{epsdelta})).

The speed of sound is defined as the coefficient that relates pressure perturbations and energy--density perturbations within the rest frame (which in the perfect-fluid case is $v_{x}=0$).
In the presence of a heat flux with
scalar potential $q_{x}$, a more appropriate definition is
\begin{equation}\label{ceps}
\hat{p}^{\varepsilon}_{x} = c_{\varepsilon}^{2} \hat{\rho}^{\varepsilon}_{x}
\end{equation}
with the accordingly defined pressure perturbation
\begin{equation}\label{}
\hat{p}^{\varepsilon}_{x} \equiv  \hat{p}_{x} - 3H\frac{\dot{p}_{x}}{\dot{\rho}_{x}}\left[\left(1+w_{x}\right)\rho_{x}v_{x} + q_{x}\right].
\end{equation}
Then, the total (isotropic) pressure perturbation on subhorizon scales $k^{2}\gg a^{2}H^{2}$ are given by
\begin{equation}\label{pcfin}
\frac{\hat{p}^{\varepsilon}}{\rho} \equiv \frac{\hat{p}^{c}}{\rho} - \frac{\dot{p}}{\dot{\rho}}l= c_{\varepsilon}^{2}\frac{\rho_{x} + p_{x}}{\rho + p}\varepsilon + c_{\varepsilon}^{2}\Omega_{m}S_{m}\qquad\qquad \left(\frac{a^{2}H^{2}}{k^{2}} \ll 1\right).
\end{equation}
Obviously, the pressure perturbations couple the dynamics of $\varepsilon$ to that of the relative perturbations $S_{m}$.
The entire dynamics is given by the coupled system of equations for $\varepsilon$ and $S_{m}$ in which there appear the so far undetermined  quantities $l$ and $\Pi$.
The simplest and most direct way to obtain a closed system consists of requiring $l$ and $\Pi$ to be linear combinations of $\varepsilon$ and $S_{m}$.
This is achieved by a general ansatz for $l$,
\begin{equation}\label{ansatzl}
l=\frac{3Hq}{\rho} =\alpha \varepsilon + \beta S_{m},
\end{equation}
where $\alpha$ and $\beta$ are phenomenological coefficients, as well as by a corresponding ansatz for $\Pi$,
\begin{equation}\label{ansatze}
\frac{H^{2}\Pi}{\rho} = \mu \varepsilon + \nu S_{m},
\end{equation}
with phenomenological coefficients $\mu$ and $\nu$.
Notice that relations (\ref{ansatzl}) and (\ref{ansatze}) are constructed to parallel the expression (\ref{pcfin}) for the isotropic pressure.
The coefficients $\alpha$ and $\beta$ quantify the r\^{o}le of the heat flux on the perturbation dynamics, the coefficients $\mu$ and $\nu$ determine the influence of the anisotropic pressure.
As the sound speed parameter $c_{x}$, the coefficients $\alpha$, $\beta$, $\mu$ and $\nu$ should be calculable from  an underlying microscopic theory. Here, these coefficients are entirely phenomenological.
The perfect-fluid case is recovered for $\alpha= \beta =\mu= \nu=0$.
Phenomenological relations between anisotropic stresses and energy--density perturbations in different contexts can be found, e.g., in \cite{kunzsapone07,cardona,blas}.
A different parametrization of perturbations via equations of state was  put forward in  \cite{battye,battye13,battye16} on the basis of a  general scalar-field Lagrangian.
Eliminating internal degrees of freedom, these authors introduced equations of state for the
entropy perturbation and the anisotropic stress in terms of perturbations of the density, the velocity and
the metric perturbations to obtain closed perturbation equations.
Their basic variables are different from ours. The entropy perturbation, e.g., is one of the two basic dynamical quantities in our context and neither velocity components nor metric functions do appear explicitly
in the system of equations.
An imperfect fluid description of scalar-tensor theories has been recently performed also in \cite{faraoni18}.

\subsection{Coupled System of Equations for $\varepsilon$ and $S_{m}$}

With relations (\ref{ansatzl}) and (\ref{ansatze}) together with Formula (\ref{pcfin}), the set of equations (\ref{epsprpr}) and (\ref{sprpr}) becomes a closed system for $\varepsilon$ and $S_{m}$.
The result, after transforming to the $k$-space ($\Delta \rightarrow -k^{2}$), is
\begin{eqnarray}\label{fineps}
&&\varepsilon^{\prime\prime} + \left(\frac{3}{2} - \frac{15}{2}\frac{p}{\rho} + 3\frac{p^{\prime}}{\rho^{\prime}} + A_{1}\right)\frac{\varepsilon^{\prime}}{a}
 - \left[\frac{3}{2} + 12\frac{p}{\rho} - \frac{9}{2}\frac{p^{2}}{\rho^{2}} - 9\frac{p^{\prime}}{\rho^{\prime}}
 - \frac{k^{2}}{a^2 H^{2}}c_{\varepsilon}^{2}\Omega_{x}
 \frac{1+w_{x}}{1+w}
  - B_{1}\right]\frac{\varepsilon}{a^{2}}\nonumber\\
&&\qquad\qquad \qquad\qquad + A_{2}\frac{S_{m}^{\prime}}{a} + \left(\frac{k^{2}}{a^{2}H^{2}}c_{\varepsilon}^{2}\Omega_{m} +B_{2}\right)\frac{S_{m}}{a^{2}}
=0,
\end{eqnarray}
where
\begin{equation}\label{}
w = \frac{p}{\rho} = \frac{p_{x}}{\rho} = w_{x}\Omega_{x},
\end{equation}
\begin{equation}\label{}
A_{1} \equiv 2\frac{k^{2}}{a^2 H^{2}} \mu - \frac{3}{2}\left(1+\frac{p}{\rho}\right)\alpha,
\end{equation}
\begin{equation}\label{}
A_{2} \equiv 2\frac{k^{2}}{a^2 H^{2}} \nu - \frac{3}{2}\left(1+\frac{p}{\rho}\right)\beta,
\end{equation}
\begin{equation}\label{}
B_{1} \equiv \left[3\left(1-\frac{p}{\rho}+2\frac{p^{\prime}}{\rho^{\prime}}\right)\frac{k^{2}}{a^{2}H^{2}}
    -\frac{2}{3} \frac{k^{4}}{a^{4}H^{4}}
    \right]\mu - \frac{9}{4}\left(1+\frac{p}{\rho}\right)\left(1-\frac{p}{\rho}\right)\alpha,
\end{equation}
\begin{equation}\label{}
B_{2} \equiv \left[3\left(1-\frac{p}{\rho}+2\frac{p^{\prime}}{\rho^{\prime}}\right)\frac{k^{2}}{a^{2}H^{2}}
    -\frac{2}{3} \frac{k^{4}}{a^{4}H^{4}}
    \right]\nu - \frac{9}{4}\left(1+\frac{p}{\rho}\right)\left(1-\frac{p}{\rho}\right)\beta
\end{equation}
and
\begin{equation}
S_{m}^{\prime\prime} +  \left[\frac{3}{2}\left(1-\frac{p}{\rho}\right) +\frac{3c_{\varepsilon}^{2}}{1+w}\Omega_{m}\right]\frac{S_{m}^{\prime}}{a}
+ M_{1}(k,a) \frac{S_{m}}{a^{2}} + E_{1}(k,a)\frac{\varepsilon^{\prime}}{a} + E_{2}(k,a)\frac{\varepsilon}{a^{2}} =0,
\label{sprpreps}
\end{equation}
with
\begin{equation}\label{}
M_{1}(k,a) \equiv \frac{1}{1+w}\left[\frac{k^{2}}{a^{2}H^{2}}c_{\varepsilon}^{2}\Omega_{m}
+3c_{\varepsilon}^{2} \left(3\frac{p}{\rho} +\left(\frac{1}{2}+ 3\frac{p^{\prime}}{\rho^{\prime}} - \frac{9}{2}\frac{p}{\rho}\right)\Omega_{m}\right)
-\frac{2}{3}\frac{k^{4}}{a^{4}H^{4}}\nu\right],
\end{equation}
\begin{equation}\label{}
E_{1}(k,a) = \frac{3}{1+w}\left(c_{\varepsilon}^{2} - \frac{p_{x}^{\prime}}{\rho_{x}^{\prime}}\right)\frac{1+w_{x}}{1+w}\Omega_{x}
\end{equation}
and
\begin{eqnarray}\label{}
E_{2}(k,a)&\equiv& \frac{1}{1+w}\left\{\left[\frac{k^{2}}{a^{2}H^{2}}c_{\varepsilon}^{2}
+3\left(\left(c_{\varepsilon}^{2} - \frac{p_{x}^{\prime}}{\rho_{x}^{\prime}}\right)\left(\frac{1}{2} + 3\frac{p^{\prime}}{\rho^{\prime}}
- \frac{9}{2}\frac{p}{\rho} -\frac{3\Omega_{m}}{1+w}\frac{p_x^{\prime}}{\rho_x^{\prime}}\right) - a \left(\frac{p_{x}^{\prime}}{\rho_{x}^{\prime}}\right)^{\prime}\right)
\right]\frac{1+w_{x}}{1+w}\Omega_{x}\right.\nonumber\\
&&\left.\qquad\qquad - \frac{2}{3}\frac{k^{4}}{a^{4}H^{4}}\mu\right\}.
\end{eqnarray}
The system (\ref{fineps}) and (\ref{sprpreps}) describes the entire linear cosmological perturbation dynamics. It represents the central set of equations for our analysis.
As we shall demonstrate in the following, its solution determines also the gravitational slip and the growth rate of matter fluctuations.

\subsection{Anisotropic Pressure and Gravitational Potentials}

The spatial part of Einstein's equation can be written as
\begin{equation}\label{Eab}
\hat{G}_{b}^{a}-\frac{1}{3}\delta^{a}_{b}\hat{G}_{m}^{m} = 8\pi G \left(\hat{T}^{a}_{b} - \frac{1}{3}\delta_{ab}\hat{T}^{m}_{m}\right),
\end{equation}
where the left-hand side is determined by the difference $\psi - \phi$,
\begin{equation}\label{}
\hat{G}_{b}^{a}-\frac{1}{3}\delta^{a}_{b}\hat{G}_{m}^{m} = \frac{1}{a^{2}}\left(\partial_{a}\partial_{b} - \frac{1}{3}\delta_{ab}\Delta\right)\left(\psi - \phi\right),
\end{equation}
while the right-hand side of Equation~(\ref{Eab}) is related to the anisotropic pressure,
\begin{equation}\label{}
\hat{T}^{a}_{b} - \frac{1}{3}\delta_{ab}\hat{T}^{m}_{m} = \Pi^{a}_{b} = \frac{1}{a^{2}}\left(\partial_{a}\partial_{b} - \frac{1}{3}\delta_{ab}\Delta\right)\Pi.
\end{equation}
It follows that
\begin{equation}\label{psi-phi=}
\psi - \phi = 8\pi G \Pi,
\end{equation}
the difference $\psi - \phi$ is directly proportional to the anisotropic pressure. Notice that, in the present gauge, the quantities $\psi$ and $\phi$ coincide with the Bardeen potentials
$\Psi$ and $\Phi$, respectively.
The relation between the comoving total density perturbation $\varepsilon$ and the gravitational potential $\psi$ is
given by Equation~(\ref{poissoneps}).
Combining Equation~(\ref{poissoneps}) with Equations~(\ref{ansatze}) and (\ref{psi-phi=}), we obtain the relation
\begin{equation}\label{phipsiS}
\phi = \left(1+ 2\mu \frac{k^{2}}{a^{2}H^{2}}\right)\psi - 3\nu S_{m}
\end{equation}
between $\psi$ and $\phi$. In the limit $\mu =\nu =0$, equivalent to a vanishing anisotropic pressure, we recover
$\psi = \phi$. In the large-scale limit $\frac{k^{2}}{a^{2}H^{2}}\ll 1$ and assuming the $S_{m}$ contribution to be small,
one has $\psi \approx \phi$ as well, but, on smaller scales, both potentials may be different.
Using Equation~(\ref{poissoneps}) to write relation~(\ref{phipsiS}) in the form
\begin{equation}\label{phipsifin}
\phi = \left[1+ 2 \frac{k^{2}}{a^{2}H^{2}}\left(\mu +\nu\frac{S_{m}}{\varepsilon}\right)\right]\psi,
\end{equation}
it is obvious that knowledge of the relation between $\psi$ and $\phi$ requires the solution of the entire coupled system of $\varepsilon$ and $S_{m}$.

\subsection{Matter Perturbations}
The final aim is to  find the matter perturbations from the solution of the system for $\varepsilon$ and $S_{m}$.
From the definition of $S_{m}$ (cf. Equation~(\ref{defSm})) together with $\delta^{c} = \varepsilon + l$, it follows that the matter perturbations are determined by
\begin{equation}\label{dml}
\delta_{m}^{c} = \frac{\varepsilon}{1+w} + \frac{l}{1+w} - S_{m}.
\end{equation}
These are the perturbations with respect to the rest frame of the cosmic fluid as a whole, characterized by the quantity $v$. It is desirable, however, to calculate the
matter-density perturbations in the matter rest frame, characterized by the  matter velocity potential $v_{m}$. Denoting this perturbation by $\delta_{m}^{c_{m}}$, the relation between both quantities is
\begin{equation}\label{deltamm}
  \delta_{m}^{c_{m}} = \delta_{m} - \Theta v_{m} = \delta_{m}^{c} + \Theta\left(v-v_{m}\right).
\end{equation}
Restricting ourselves to sub-horizon scales again, the matter density perturbations in the matter rest frame are
\begin{equation}\label{deltamfin}
\delta_{m}^{c_{m}} =\frac{\varepsilon}{1+w_{x}\Omega_{x}}  - S_{m}\qquad\qquad \left(\frac{a^{2}H^{2}}{k^{2}} \ll 1\right).
\end{equation}
These  matter perturbations are
directly given by the solution of the coupled system (\ref{fineps}) and (\ref{sprpreps}) for $\varepsilon$ and $S_{m}$, respectively.

The entire setup so far is completely general and does not use any specific model for the $x$ component.
It is also valid for any homogeneous and isotropic background.
Since any generalized or modified (compared with GR) gravitational theory can formally be rewritten as an effective Einsteinian theory, our formalism is expected to be applicable  for a broad range of models.
In the following, we apply this general scheme to a previously established scalar-tensor extension of the $\Lambda$CDM model, called $e_{\Phi}\Lambda$CDM model, which provides us with a non-standard background dynamics.
This completes a preliminary study in \cite{WHW2}, where the matter growth rate was obtained in a simplified manner on the basis of a rough approximation without solving the full dynamics given by Equations~(\ref{fineps}) and (\ref{sprpreps}) and without including the heat flux.

\section{$e_{\Phi}\Lambda$CDM Cosmology}
\label{ephilambda}
\subsection{Jordan--Brans--Dicke Theory}

Our example originates from Jordan--Brans--Dicke (JBD) scalar-tensor theory \cite{jordan,BD,B}, which was inspired by earlier ideas of Mach.
Generally,
the gravitational interaction in scalar-tensor theories is mediated by a scalar field in addition to the GR-type interaction through the metric tensor.
Our aim is to find an equivalent GR description of such extended gravitational theory
by mapping the additional (compared with GR) geometric degrees of freedom onto an effective  fluid component. The energy-momentum tensor of this effective fluid will, in general, be of the structure of an imperfect fluid, i.e., it gives rise to effective anisotropic stresses and energy fluxes which are absent in a perfect-fluid based GR description.
With an effective imperfect fluid description at hand, the perturbation analysis of the previous sections applies straightforwardly. In particular, it will be possible to calculate the growth rate of matter perturbations. What is needed, however, is a solution for the homogeneous and isotropic background which determines the coefficients in the set of first-order perturbation equations. The background is no longer that of the standard $\Lambda$CDM model, but it has to be the background of the extended gravitational theory. The first part of this section deals with the issue of how to obtain a suitable solution for a homogeneous and isotropic dynamics that deviates from the $\Lambda$CDM model.

Starting from a JBD type action (see, e.g., \cite{abean,scalten,clifton,chibayam})
\begin{equation}
\label{action1}
   S(g_{\mu\nu},\Phi)= \frac{1}{2\kappa^{2}}\int d^{4}x\sqrt{-g}\left[\Phi R - \frac{\omega(\Phi)}{\Phi}\left(\nabla \Phi\right)^{2} - U(\Phi)\right] + S_{m}\left(g_{\mu\nu}\right),
\end{equation}
where $S_{m}$ is the matter part,
the Jordan-frame gravitational field equations for scalar-tensor theories are
\begin{eqnarray}
\label{eqJ}
\Phi\left(R_{\mu\nu}- \frac{1}{2}g_{\mu\nu}R\right) &=& \kappa^{2}T_{(m)\mu\nu}
\nonumber\\
&& + \frac{\omega(\Phi)}{\Phi}
\left(\partial_{\mu}\Phi\partial_{\nu}\Phi
- \frac{1}{2}g_{\mu\nu}\left(\nabla\Phi\right)^{2}\right)
+ \nabla_{\mu}\nabla_{\nu}\Phi- g_{\mu\nu}\Box\Phi
- \frac{1}{2}g_{\mu\nu}U,
\end{eqnarray}
with $\kappa^{2} \equiv 8\pi G$ and
\begin{equation}\label{boxphi}
\Box \Phi = \frac{1}{2\omega(\Phi) + 3}\left(\kappa^{2}T - \frac{d\omega(\Phi)}{d\Phi}\left(\nabla\Phi\right)^{2} + \Phi\frac{dU}{d\Phi} - 2U\right),
\end{equation}
where $T$ is the trace of the matter energy-momentum tensor
\begin{equation}\label{}
T_{(m)\mu\nu} = - \frac{2}{\sqrt{-g}}\frac{\delta S_{m}}{\delta g^{\mu\nu}}.
\end{equation}
The choice of the symbol $T_{(m)\mu\nu}$ indicates that we shall identify this quantity with the matter energy-momentum tensor of Section \ref{EMT}.
In order to relate the entire formalism to the quantities of Section \ref{EMT}, we introduce a total effective energy-momentum tensor (cf.~Equation~(\ref{Ttot}))
$T_{\mu\nu} = T_{(m)\mu\nu} + T_{(x)\mu\nu}$,
where $T_{(x)\mu\nu}$ is an effective part that describes geometric ``matter" and has the structure
\begin{equation}\label{TxJ}
T_{(x)\mu\nu} \equiv \left(\frac{1}{\Phi}-1\right)T_{(m)\mu\nu} + \frac{1}{\kappa^{2}\Phi}
\left[\frac{\omega(\Phi)}{\Phi}
\left(\partial_{\mu}\Phi\partial_{\nu}\Phi
- \frac{1}{2}g_{\mu\nu}\left(\nabla\Phi\right)^{2}\right)
+ \nabla_{\mu}\nabla_{\nu}\Phi- g_{\mu\nu}\Box\Phi
- \frac{1}{2}g_{\mu\nu}U\right].
\end{equation}
Again, we have used the symbol $T_{(x)\mu\nu}$, introduced in Section \ref{EMT}, to indicate identification
with expression  (\ref{TxJ}). The same is true for the sum $T_{\mu\nu} = T_{(m)\mu\nu} + T_{(x)\mu\nu}$.
With this definition, the field equation (\ref{eqJ}) acquires
the Einsteinian form,
\begin{equation}\label{gr}
R_{\mu\nu}- \frac{1}{2}g_{\mu\nu}R = \kappa^{2}T_{\mu\nu}.
\end{equation}
Having written JBD theory as an effective GR theory allows us to use techniques developed within the latter to make statements concerning the former as well.
Our focus will be here on the background dynamics.
In a spatially flat homogeneous and isotropic model, one may associate a Hubble rate to the scalar-tensor dynamics \cite{chibayam}:
\begin{equation}\label{H2J}
H^{2} = \frac{\kappa^{2}}{3}\frac{\rho_{m}}{\Phi} + \frac{1}{3\Phi}
\left[\frac{1}{2}\frac{\omega(\Phi)}{\Phi}
  \left(\frac{\partial\Phi}{\partial t}\right)^{2}
  -3H\frac{\partial\Phi}{\partial t}
  + \frac{1}{2}U\right],
\end{equation}
where $H = \frac{1}{a}\frac{da}{dt}$ is the Hubble rate of the Jordan frame and
$a$ is the Jordan-frame scale factor. Furthermore, assuming the matter component to be pressureless, one has
\begin{equation}\label{}
\frac{d H}{d t} = - \frac{\kappa^{2}}{2} \frac{\rho_{m}}{\Phi}
- \frac{1}{2\Phi}\left[\frac{\omega(\Phi)}{\Phi}
  \left(\frac{\partial\Phi}{\partial t}\right)^{2}
  -H\frac{\partial\Phi}{\partial t}
  + \frac{d^{2}\Phi}{dt^{2}}\right],
\end{equation}
and
\begin{equation}\label{equPhigen}
\frac{d^{2}\Phi}{dt^{2}} + 3 H\frac{d\Phi}{dt} = \frac{1}{2\omega + 3}\left(\kappa^{2}\rho_{m} - \frac{d\omega}{d\Phi}\left(\frac{d\Phi}{dt}\right)^{2} - \Phi\frac{dU}{d\Phi} + 2U\right),
\end{equation}
as well as the matter conservation (\ref{mattercons}).

The structure of Equation~(\ref{H2J}) suggests the definition
\begin{equation}\label{rhoxPhi}
\rho_{x_{(\Phi)}} = \frac{\rho_{m}}{3}\left(\frac{1}{\Phi}-1\right) + \frac{1}{3\kappa^{2}\Phi}
\left[\frac{1}{2}\frac{\omega(\Phi)}{\Phi}
  \left(\frac{\partial\Phi}{\partial t}\right)^{2}
  -3H\frac{\partial\Phi}{\partial t}
  + \frac{1}{2}U\right]
\end{equation}
such that Equation~(\ref{H2J}) takes the form of an effective Friedmann equation $3H^{2} = \kappa^{2}\left(\rho_{m} + \rho_{x_{(\Phi)}}\right)$.
Obviously, $\rho_{x_{(\Phi)}}$ is determined by the dynamics of the scalar field, this is indicated by the additional subscript $(\Phi)$.

\subsection{JBD Inspired Effective Background Model}
\label{background}
Here, we recall the basic elements of the previously established homogeneous and isotropic scalar-tensor extension of the $\Lambda$CDM model \cite{WHW}.
Its main characteristic is an explicit analytical expression for the Hubble rate in which deviations from the standard model are described by a single constant parameter which also governs the effective scalar-field dynamics. The background relations are found through a specific solution of the effective fluid dynamics in the Einstein frame which subsequently is converted into the Jordan frame via a conformal transformation.
To make our presentation self-contained, we summarize the main steps of the corresponding derivation.
The starting point for the background analysis is the Einstein frame. The reason for this apparent detour is that it is the Einstein frame in which it is possible to obtain a solution of the dynamics.
\subsubsection{General Einstein-Frame Dynamics}
Generally, the Einstein-frame dynamics with a metric $\tilde{g}_{\mu\nu}$ follows from the (so far considered) Jordan-frame dynamics with a metric $g_{\mu\nu}$ through the conformal transformation
\begin{equation}
g_{\mu\nu} = \frac{1}{\Phi}\tilde{g}_{\mu\nu} = e^{2\,b(\varphi)}\,\tilde{g}_{\mu\nu}
 \ \label{gtil}
\end{equation}
together with a redefinition of the potential term,
\begin{equation}\label{VU}
V(\varphi) = \frac{U(\Phi)}{2\kappa^{2}\Phi^{2}}.
\end{equation}
Moreover, one has
\begin{equation}\label{Phivarphi}
\frac{1}{4\Phi^{2}}\left(\frac{d\Phi}{d\varphi}\right)^{2} = \left(\frac{d b}{d\varphi}\right)^{2}
= \frac{\kappa^{2}}{4\omega(\Phi)+6}.
\end{equation}
For an RW metric and assuming $u^{\mu}_{(m)}= u^{\mu}$, the general relations simplify considerably.
With the Einstein-frame scale factor $\tilde{a}$ and the Einstein-frame time coordinate $\tilde{t}$ the Einstein-frame Hubble rate
 $\tilde{H} = \frac{1}{\tilde{a}}\frac{d\tilde{a}}{d\tilde{t}}$ becomes
\begin{equation}
\tilde{H}^{2} = \frac{\kappa^{2}}{3}\left[\tilde{\rho}_{m} + \frac{1}{2}\left(\frac{d\varphi}{d\tilde{t}}\right)^{2} + \tilde{V}\right],
\label{tH2}
\end{equation}
where $\tilde{\rho}_{m}$ is the matter-energy density in the Einstein frame.
Quantities with a tilde have their meaning in the Einstein frame, those without a tilde refer to the Jordan frame.
Different from the Jordan-frame dynamics, the matter part is not separately conserved but obeys the equation
\begin{equation}
\frac{d\tilde{\rho}_{m}}{d\tilde{t}} + 3\tilde{H} \tilde{\rho}_{m}  =
\frac{d\varphi}{d\tilde{t}}\frac{db}{d\varphi}
\tilde{\rho}_{m}
\label{drmn}
\end{equation}
through which it couples to the following scalar-field dynamics:
\begin{equation}
 \frac{d^{2}\varphi}{d\tilde{t}^{2}} + 3\tilde{H}\frac{d\varphi}{d\tilde{t}} + \tilde{V}_{,\varphi} =
-\frac{db}{d\varphi}
\tilde{\rho}_{m}.
\label{ddvphi}
\end{equation}
Here, $\tilde{V}_{,\varphi} = \frac{\partial \tilde{V}}{\partial{\varphi}}$.
The time coordinate $t$, the scale factor $a$ and the matter energy density $\rho_{m}$  of the Jordan frame  are related to their Einstein-frame counterparts by
\begin{equation}
dt =
e^{b} d \tilde{t},\ \qquad a =
e^{b}\tilde{a}\qquad \mathrm{and} \qquad \rho_{m} = e^{-4 b}\,\tilde{\rho}_{m},\label{ttilb}
\end{equation}
respectively.

\subsubsection{Interacting Fluid Approach in Einstein-Frame Dynamics}

We associate an effective energy density $\tilde{\rho}_{\varphi}$ and an effective pressure $\tilde{p}_{\varphi}$ to the scalar field by
\begin{equation}
\tilde{\rho}_{\varphi} = \frac{1}{2}\left(\frac{d\varphi}{d\tilde{t}}\right)^{2}+ \tilde{V}\qquad \mathrm{and} \qquad \tilde{p}_{\varphi} =
\frac{1}{2}\left(\frac{d\varphi}{d\tilde{t}}\right)^{2} - \tilde{V},
\label{rhophi}
\end{equation}
respectively.
Equation (\ref{ddvphi}) then takes the form
\begin{equation}
\frac{d\tilde{\rho}_{\varphi}}{d\tilde{t}}  + 3\tilde{H} \left(1 +
\tilde{w}\right)\tilde{\rho}_{\varphi} = - Q
\equiv - \frac{d\varphi}{d\tilde{t}}\frac{d b}{d\varphi}
\tilde{\rho}_{m},
\label{drp+b}
\end{equation}
where $\tilde{w} = \frac{\tilde{p}_{\varphi}}{\tilde{\rho}_{\varphi}}$ is the Einstein-frame EoS parameter for the scalar field.
Equation (\ref{drmn}) can be written as
\begin{equation}
\tilde{\rho}_{m} = \tilde{\rho}_{m0}\tilde{a}^{-3}f(\tilde{a}) \quad\Rightarrow \quad
\frac{d\tilde{\rho}_{m}}{d\tilde{t}}  + 3\tilde{H} \tilde{\rho}_{m} = \frac{\tilde{\rho}_{m}}{f}
\frac{d f}{d\tilde{t}}, \quad
\Rightarrow\quad f = e^{b(\varphi)},
 \ \label{rhof}
\end{equation}
where the function $f$ encodes the effects of an interaction between matter and (Einstein-frame) scalar field.
Assuming a power-law behavior of the interaction function $f\left(\tilde{a}\right)$,
\begin{equation}\label{pansatz}
f\left(\tilde{a}\right) = \tilde{a}^{3m},
\end{equation}
the explicit solution of Equation~(\ref{drp+b}) for a constant EoS parameter then is
\begin{equation}
\tilde{\rho}_{\varphi} = \tilde{\rho}_{\varphi 0}\tilde{a}^{-3\left(1+\tilde{w}\right)} - \tilde\rho_{m_0}\,\tilde{a}^{-3\left(1+\tilde{w}\right)}\, \int_{\tilde{a}_{0}}^{\tilde{a}} d\tilde{a}
\frac{d f}{d \tilde{a}}\,\tilde{a}^{3\tilde{w}}.
\label{rx3}
\end{equation}
To obtain the explicit solutions (\ref{rhof}) with (\ref{pansatz}) and (\ref{rx3}) was the main motivation for considering the Einstein frame.
This implies the expression
\begin{equation}\label{Qm}
Q = 3m\tilde{H}\tilde{\rho}_{m}
\end{equation}
for
the interaction term $Q$, defined in (\ref{drp+b}).
In the
simple case of a linear dependence
\begin{equation}\label{blin}
b=b(\varphi) = K\varphi, \qquad K = \sqrt{\frac{\kappa^{2}}{4\omega +6}},
\end{equation}
it follows that \cite{WHW}
\begin{equation}\label{Phivarphi+}
\Phi = e^{-2K\varphi} = e^{-2b} =\tilde{a}^{-6m} = a^{-\frac{6m}{1+3m}}.
\end{equation}
With Formula~(\ref{Phivarphi+}), we found an explicit expression for the scalar-field variable without having solved the basic scalar-field equation.
We have solved the system of energy-balance equations (\ref{drmn}) and (\ref{drp+b}) under the assumptions (\ref{pansatz}) and (\ref{blin}).
This solution implies an explicit scale-factor dependence of $\Phi$ (or $\varphi$) which does not necessarily have to be a solution of the original scalar-field equation (\ref{boxphi}).
Instead, it obeys an alternative effective second-order equation with an alternative effective potential that does not coincide with $U$ \cite{WHW}.
The point here is that the dynamics on the level of the fluid energy densities do not require the exact solution of the scalar-field equation (\ref{boxphi}).
On the other hand, the specific features of our fluid dynamics imply the existence of the effective scalar field $\Phi$ of the form of  (\ref{Phivarphi+}).
Under such condition, the dynamics for $\tilde{w}=-1$ are explicitly solved, which results in the expression \cite{WHW}
\begin{equation}\label{tilh2+}
\frac{\tilde{H}^{2}}{\tilde{H}^{2}_{0}} = e^{K\varphi}\frac{\tilde{\Omega}_{m0}\tilde{a}^{-3}}{1 - m}
+ 1  - \frac{\tilde{\Omega}_{m0}}{1 - m},\qquad \tilde{\Omega}_{m0} = \frac{8\pi G\tilde{\rho}_{m0}}{3\tilde{H}_{0}^{2}}
\end{equation}
for the Einstein-frame Hubble rate.
Having used the Einstein frame to explicitly solve the background dynamics, we now turn back to the Jordan-frame dynamics.

\subsubsection{Effective Hubble Rate}
The transformation to the
Jordan-frame Hubble rate via
\begin{equation}\label{}
H= e^{-b}\left(1+3m\right)\tilde{H}
\end{equation}
leads to the explicit effective Hubble rate \cite{WHW}
\begin{equation}\label{H2Phi}
\frac{H^{2}}{H_{0}^{2}} =
\frac{A\Omega_{m0}a^{-3}}{\Phi}
+ \left[1 - A\Omega_{m0}\right]\Phi,
\end{equation}
where
\begin{equation}\label{}
\Omega_{m0} = \frac{\tilde{\Omega}_{m0}}{\left(1+3m\right)^{2}},\qquad
A\equiv \frac{\left(1 +3m \right)^{2}}{1 - m}
\end{equation}
and $\Phi$ is explicitly given in terms of the scale factor by Relation~(\ref{Phivarphi+}).
Formula (\ref{H2Phi}) represents an explicit analytic solution for the Hubble rate of our $e_{\Phi}\Lambda$CDM model.
The scalar $\Phi$ modifies the cosmological dynamics compared
with the GR based $\Lambda$CDM model.
For $\Phi = 1$, equivalent to $m=0$, we recover the $\Lambda$CDM model.
For any $\Phi \neq 1$, equivalent to $m\neq 0$, the expression (\ref{H2Phi})  represents a testable, alternative model with  small ($|m|\ll 1$) deviations from the $\Lambda$CDM model.
It is formula (\ref{H2Phi}) which allows us to apply the general perturbation dynamics of the previous sections to a specific class of non-standard models. To motivate the result, (\ref{H2Phi}) has been the main purpose of this section.
The dependence of the Hubble rate (\ref{H2Phi}) on the scalar $\Phi$ changes the relative contributions of matter and the dark-energy (DE) equivalent compared with the $\Lambda$CDM model.
Deviations from the $\Lambda$CDM model are entirely encoded in the constant parameter $m$ with $|m|\ll 1$.
The fractional matter contribution is
\begin{equation}\label{}
\Omega_{m} = \frac{\rho_{m}}{\rho} = \frac{\Omega_{m0}a^{-3}}{A\Phi^{-1}\Omega_{m0}a^{-3} + \left[1 - A\Omega_{m0}\right]\Phi},
\end{equation}
the geometric ``matter" part contributes with $\Omega_{x} = 1 - \Omega_{m}$.
For different values of $m$, the fractional abundances $\Omega_{m}$ and $\Omega_{x}$ are visualized in Figure~\ref{figOmxn}.
Postulating a conservation equation $\dot{\rho}_{x} + 3H\left(1+w_{x}\right)\rho_{x} =0$, where $\rho_{x} \equiv \rho - \rho_{m}$ now replaces Expression~(\ref{rhoxPhi}), this corresponds to an effective, time-varying EoS parameter $w_{x}$ of the geometric DE,
\begin{equation}\label{wtil-1}
w_{x}(a) = - 1 + \frac{\frac{2m}{1+3m}\left[1 - A\Omega_{m0}\right]\Phi + \Omega_{m0}a^{-3}\left[\frac{1+m}{1+3m}A \Phi^{-1} - 1\right]}{\left[1 - A\Omega_{m0}\right]\Phi + \Omega_{m0}a^{-3}
\left[A \Phi^{-1} - 1\right]}.
\end{equation}

\begin{figure}[H]
\centering
{\includegraphics[width=0.4\textwidth]{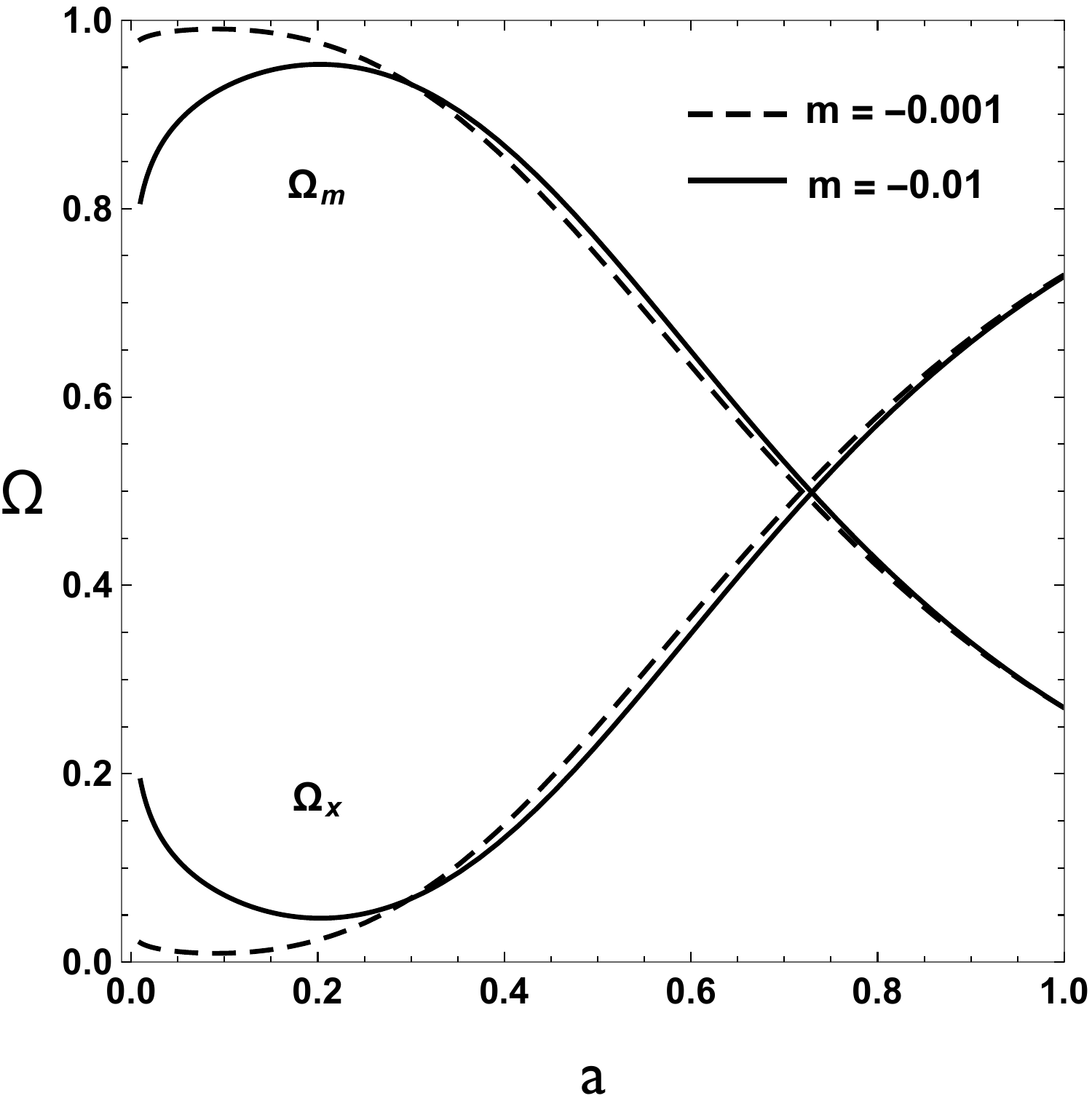}}
{\includegraphics[width=0.4\textwidth]{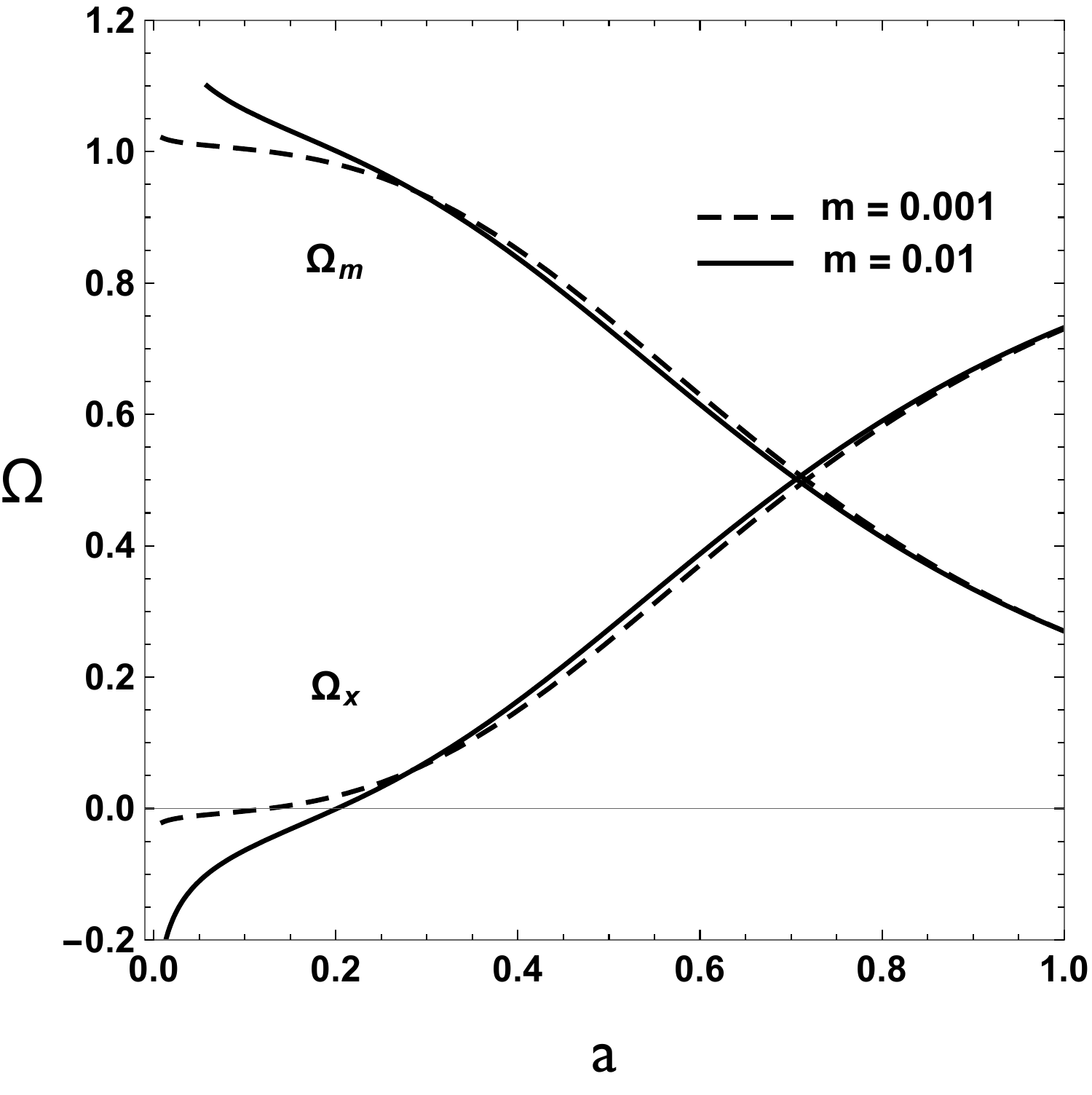}}
\label{figOmxn}	
\caption{Matter fraction $\Omega_m$  and geometric energy fraction $\Omega_x$ for negative (left) and positive
(right) values of $m$. } 
\label{figOmxn}
\end{figure}

\begin{figure}[H]
\centering
{\includegraphics[width=0.45\textwidth]{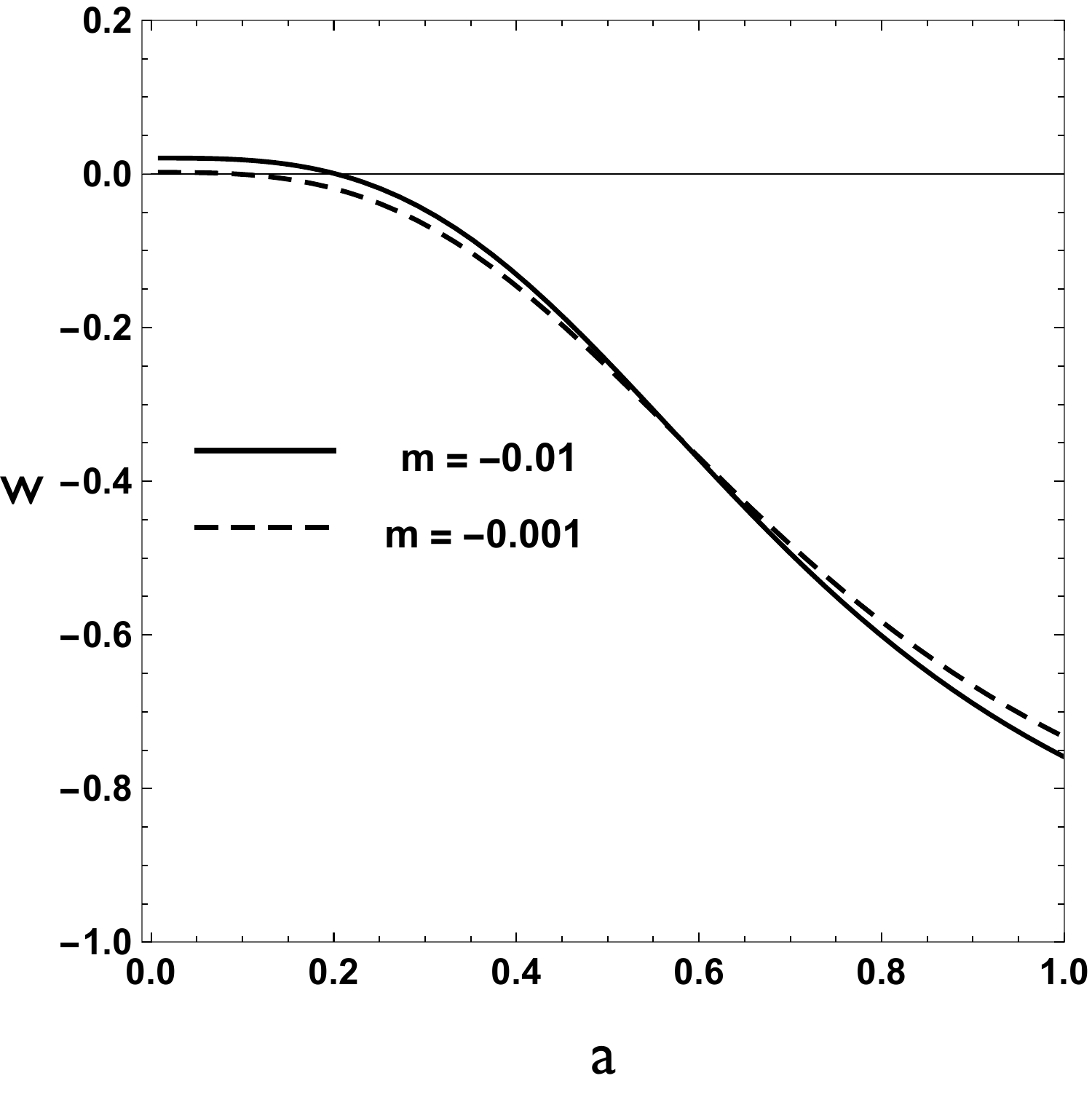}}
{\includegraphics[width=0.45\textwidth]{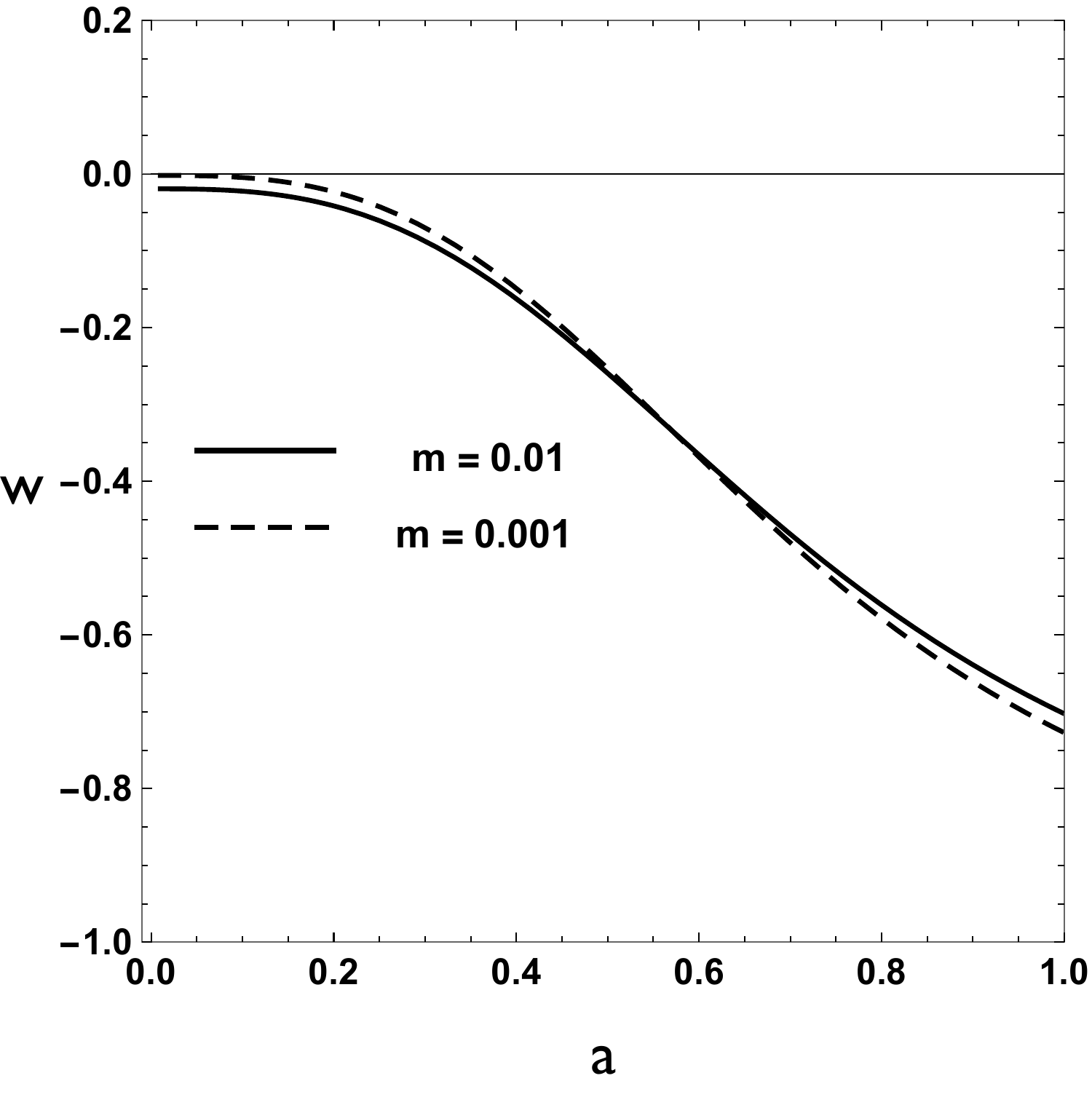}}
\caption{Total EoS
parameter $w=\frac{p}{\rho}$ for various negative (left) and positive (right) values of $m$.}
\label{figOmxp}	
\end{figure}

For $m=0$, it reduces to the $\Lambda$CDM value $w_{x} = - 1$.
At a high redshift, one has
 \begin{equation}\label{}
w_{x} \approx - 1 + \frac{\left[\frac{1+m}{1+3m}A \Phi^{-1} - 1\right]}{\left[A \Phi^{-1} - 1\right]}
\qquad\qquad (a\ll 1).
\end{equation}
 This value may be close to zero, i.e., the geometric DE may mimic dust in this limit,
but the effective energy density $\rho_{x}$ will be negative for $m>0$. It crosses $\rho_{x}=0$ in the redshift range $10\gtrsim z \gtrsim 4$ for the values of $m$ chosen in
Figure~\ref{figOmxn}.
This behavior reflects that fact that the $x$-component is very different from a conventional fluid.
The total EoS is well behaved throughout  as is demonstrated in Figure~\ref{figOmxp}.
At the present time, the effective EoS parameter is
\begin{equation}\label{}
w_{x} = - 1 + \frac{\frac{2m}{1+3m} +  3m\Omega_{m0}}{1 - \Omega_{m0}} \qquad\qquad (a=1).
\end{equation}
For small $|m|$, this remains in the vicinity of $w_{x} = - 1$.
In the far future, $w_{x}$ approaches
\begin{equation}\label{}
w_{x} \approx - 1 + \frac{2m}{1+3m}  \qquad\qquad (a\gg 1).
\end{equation}

From a statistical analysis using Supernovae data, data from the
differential age of old galaxies that have evolved passively and baryon acoustic oscillations, we found a best-fit value \cite{WHW} of
$m= 0.004^{+0.011 (1\sigma) \,\, +0.017 (2\sigma)}_{-0.011 (1\sigma)\,\, -0.017 (2\sigma)}$.
This is compatible with the $\Lambda$CDM model, but it leaves room for small deviations.
Even a very small non-vanishing value of $|m|$ will modify the standard scenario of structure formation.
The quantitative analysis to follow will rely on the use of the effective Hubble rate (\ref{H2Phi}) for the background coefficients of the perturbation equations.


\section{Growth of Matter Perturbations}
\label{data}

\begin{figure}[H]
\centering
\includegraphics[width=0.6\textwidth]{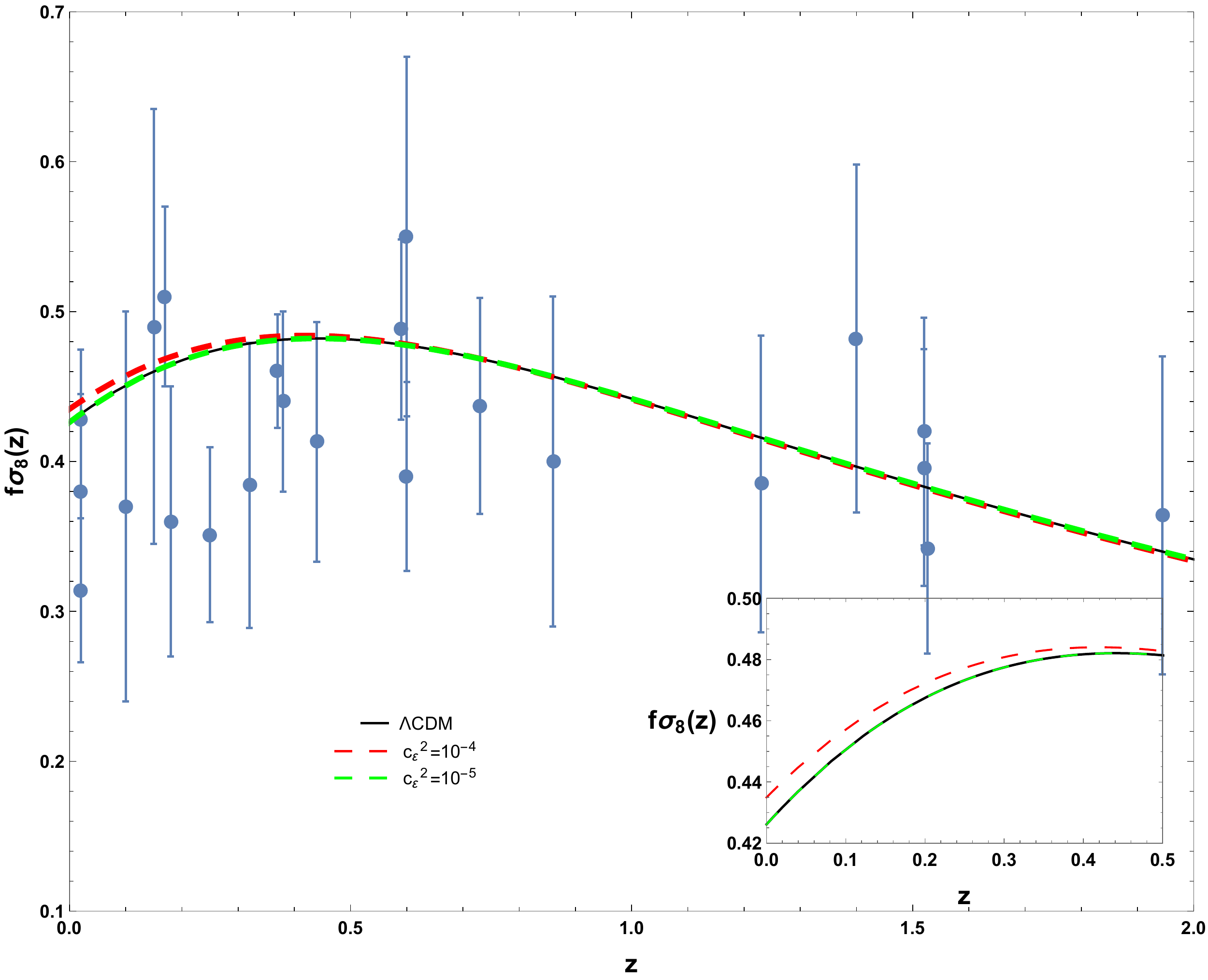}
\caption{Dependence of $f\sigma_8 (z)$ on $z$ for a non-vanishing sound-speed parameter ($m= \alpha = \beta = \mu= \nu =0$).} 
\label{deltam-1}	
\end{figure}

Now, we combine the general imperfect-fluid perturbation dynamics, established in Section \ref{Perturbation}, with the $e_{\Phi}\Lambda$CDM background model of the previous section.
The variable of interest, the fractional matter perturbation $\delta_{m}^{c_{m}}$ in Formula~(\ref{deltamfin}), is obtained via the solution of the coupled system (\ref{fineps})
and (\ref{sprpreps}) for $\varepsilon$ and $S_{m}$, respectively. Since on sub-horizon scales gauge issues are less important, we shall omit from now on the superscript $c_{m}$ and denote this quantity simply by $\delta_{m}$.
It is convenient to use also the linear growth rate $f$ defined by $f= {\rm d\, ln}\, \delta_m / {\rm d\, ln}\, a$.
In most cases, observational data are provided for the combination $f \sigma_8$ where $\sigma_8$ is the root-mean-square mass fluctuation in spheres with radius $8h^{-1}\mathrm{Mpc}$~\cite{song09}.
In the linear regime, one has  \cite{nesseris08}
\begin{eqnarray}
\sigma _8(z)=\frac{\delta_{m} (z)}{\delta (z=0)}\sigma _8(z=0)
\end{eqnarray}
and
\begin{eqnarray}
f\sigma_8 (z)= -(1+z)\frac{\sigma_8(z=0)}{\delta_{m}(z=0)}\frac{d}{dz}\delta_{m}(z).
\end{eqnarray}
For the matter distribution variance today, we assume $\sigma_8(z=0)=0.8$, which is compatible with standard fiducial cosmology as determined by the Planck satellite \cite{planck18}. This value is biased with respect to the variance in the distribution of galaxies but the combination $f \sigma_8$ is independent from the bias factor \cite{song09}.
The data of RSD measurements of $f \sigma_8$ used in our analysis are listed in Table \ref{table1}.
  In the following, we study the influence of some of the model parameters on the growth of matter perturbations.
  Differences to the $\Lambda$CDM model may already occur if there are isotropic pressure perturbations, described by a non-vanishing sound speed parameter.
  Our analysis shows that values of the order of $c_{\varepsilon}^{2} > 10^{-4}$ are necessary to produce a noticeable impact (Figure~\ref{deltam-1}).

  \begin{table}[h]
\centering
\caption{\label{table1} Data points for our analysis.
The first 18 entries represent the ``Gold'' set of \cite{nesseris17}. The entries 19 and 20 are taken from  \cite{sdssIV}  and  \cite{sdssIV2}, respectively. The last three values have been reported in \cite{sdssIV3}.}
\vspace{0.5cm}
\begin{tabular}{l c  c  c  c  c}

\hline
\bf {Index} & \bf {Data set} & $\bf {z}$ & $ f\sigma_8(z)$ & \bf {Year}  & \bf {Notes}\\ 
\hline                               
1 & 6dFGS + SnIa        &  0.02   &   0.428 $\pm$ 0.0465   &   2016   &    ($\Omega_{0m}$, $h$, $\sigma_8$)= (0.3, 0.683, 0.8)\\
2 & SnIa + IRAS         &  0.02   & 0.398 $\pm$ 0.065      & 2011     & ($\Omega_{0m}$, $\Omega_K$)=(0.3, 0)  \\
3 & 2MASS               &  0.02   & 0.314 $\pm$ 0.048      & 2010     & ($\Omega_{0m}$, $\Omega_K$)=(0.266,0)  \\
4 & SDSS-veloc          &  0.10   & 0.370 $\pm$ 0.130      & 2015     & ($\Omega_{0m}$, $\Omega_K$)=(0.3, 0)  \\
5 & SDSS-MGS            &  0.15   & 0.490 $\pm$ 0.145      & 2014     & ($\Omega_{0m}$, $h$, $\sigma_8$)= (0.31, 0.67, 0.83)  \\          
6 & 2dFGRS              &  0.17   & 0.510 $\pm$ 0.060      & 2009     & ($\Omega_{0m}$, $\Omega_K$)=(0.3, 0)  \\
7 & GAMA                &  0.18   & 0.360 $\pm$ 0.090      & 2013     & ($\Omega_{0m}$, $\Omega_K$)=(0.27, 0)  \\
8 & GAMA                &  0.38   & 0.440 $\pm$ 0.060      & 2013     &   \\
9 & SDSS-LRG-200        &  0.25   & 0.3512 $\pm$ 0.0583    & 2011     & ($\Omega_{0m}$, $\Omega_K$)=(0.25, 0)  \\
10 & SDSS-LRG-200       &  0.37   & 0.4602 $\pm$ 0.0378    & 2011     &   \\
11 & BOSS-LOWZ          &  0.32   & 0.384 $\pm$ 0.095      & 2013     & ($\Omega_{0m}$, $\Omega_K$)=(0.274, 0)  \\
12 & SDSS-CMASS         &  0.59   & 0.488 $\pm$ 0.060      & 2013     & ($\Omega_{0m}$, $h$, $\sigma_8$)= (0.307115, 0.6777, 0.8288)  \\
13 & WiggleZ            &  0.44   & 0.413 $\pm$ 0.080      & 2012     & ($\Omega_{0m}$, $h$)=(0.27, 0.71)  \\
14 & WiggleZ            &  0.60   & 0.390 $\pm$ 0.063      & 2012     &   \\
15 & WiggleZ            &  0.73   & 0.437 $\pm$ 0.072      & 2012     &   \\
16 & Vipers PDR-2       &  0.60   & 0.550 $\pm$ 0.120      & 2016     & ($\Omega_{0m}$, $\Omega_b$)=(0.3, 0.045)  \\
17 & Vipers PDR-2       &  0.86   & 0.400 $\pm$ 0.110      & 2016     &   \\
18 & FastSound          &  1.40   & 0.482 $\pm$ 0.116      & 2015     & ($\Omega_{0m}$, $\Omega_K$)=(0.270, 0)  \\
19 & SDSS-IV            &  1.52   & 0.420 $\pm$ 0.076      & 2018     & ($\Omega_{0m}$, $\Omega_bh^2$, $\sigma_8$)= (0.26479, 0.02258, 0.8)  \\
20 & SDSS-IV            &  1.52   & 0.396 $\pm$ 0.079      & 2018     & ($\Omega_{0m}$,$\Omega_bh^2$, $\sigma_8$)= (0.31, 0.022, 0.8225)  \\
21 & SDSS-IV            &  1.23   & 0.385 $\pm$ 0.099      & 2018     & ($\Omega_{0m}$, $\sigma_8$)= (0.31, 0.8)  \\
22 & SDSS-IV            &  1.526   & 0.342 $\pm$ 0.070      & 2018     &   \\
23 & SDSS-IV            &  1.944   & 0.364 $\pm$ 0.106      & 2018     &\\
\hline
\end{tabular}
\end{table}

\begin{figure}[H]
\centering
\includegraphics[width=0.6\textwidth]{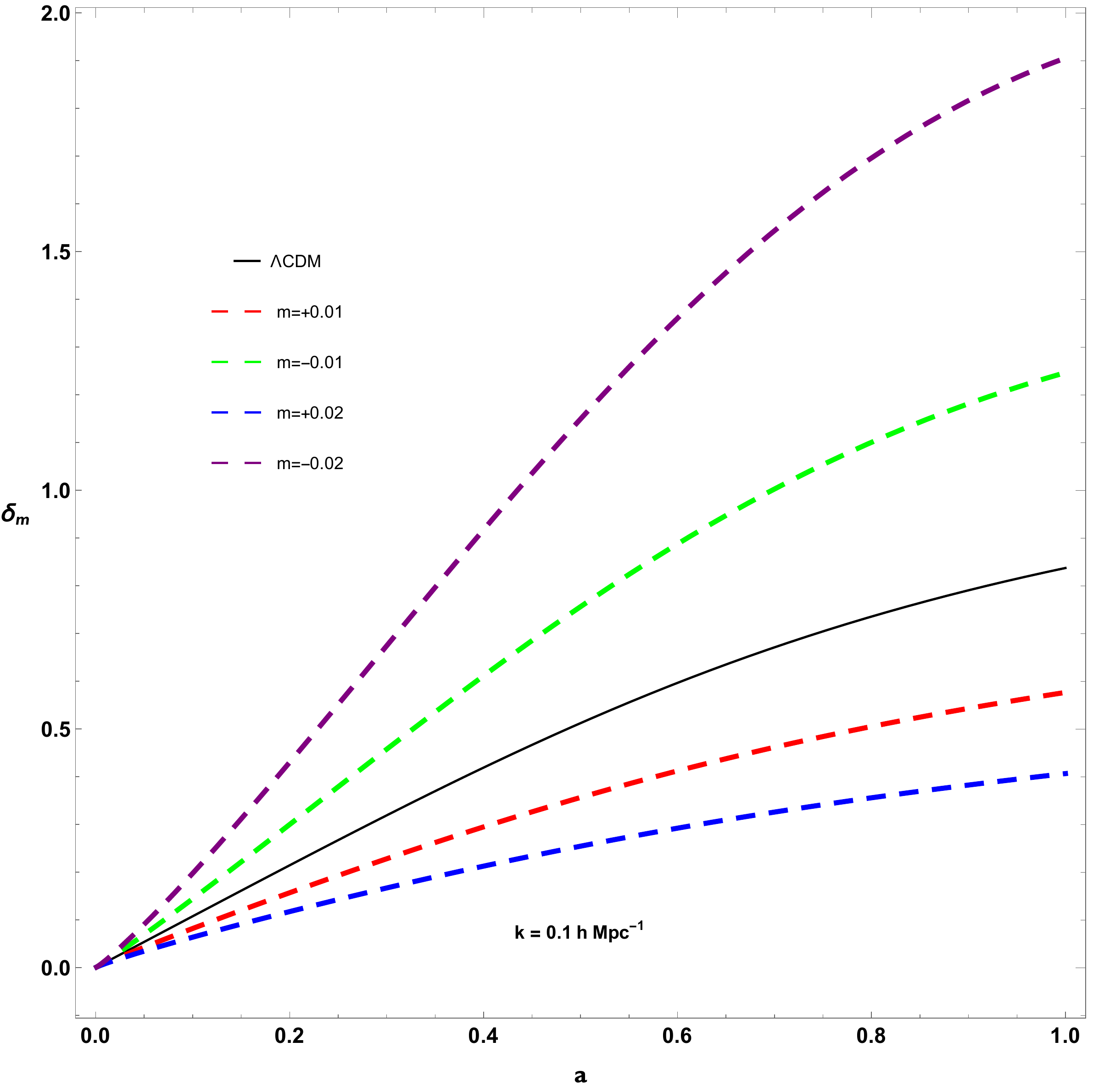}
\caption{Matter growth for various values of the parameter $m$ with $\alpha = \beta = \mu= \nu = c_{\varepsilon}^{2} =0$.}
\label{delta-m}
\end{figure}

\begin{figure}[H]
\centering
\includegraphics[width=0.6\textwidth]{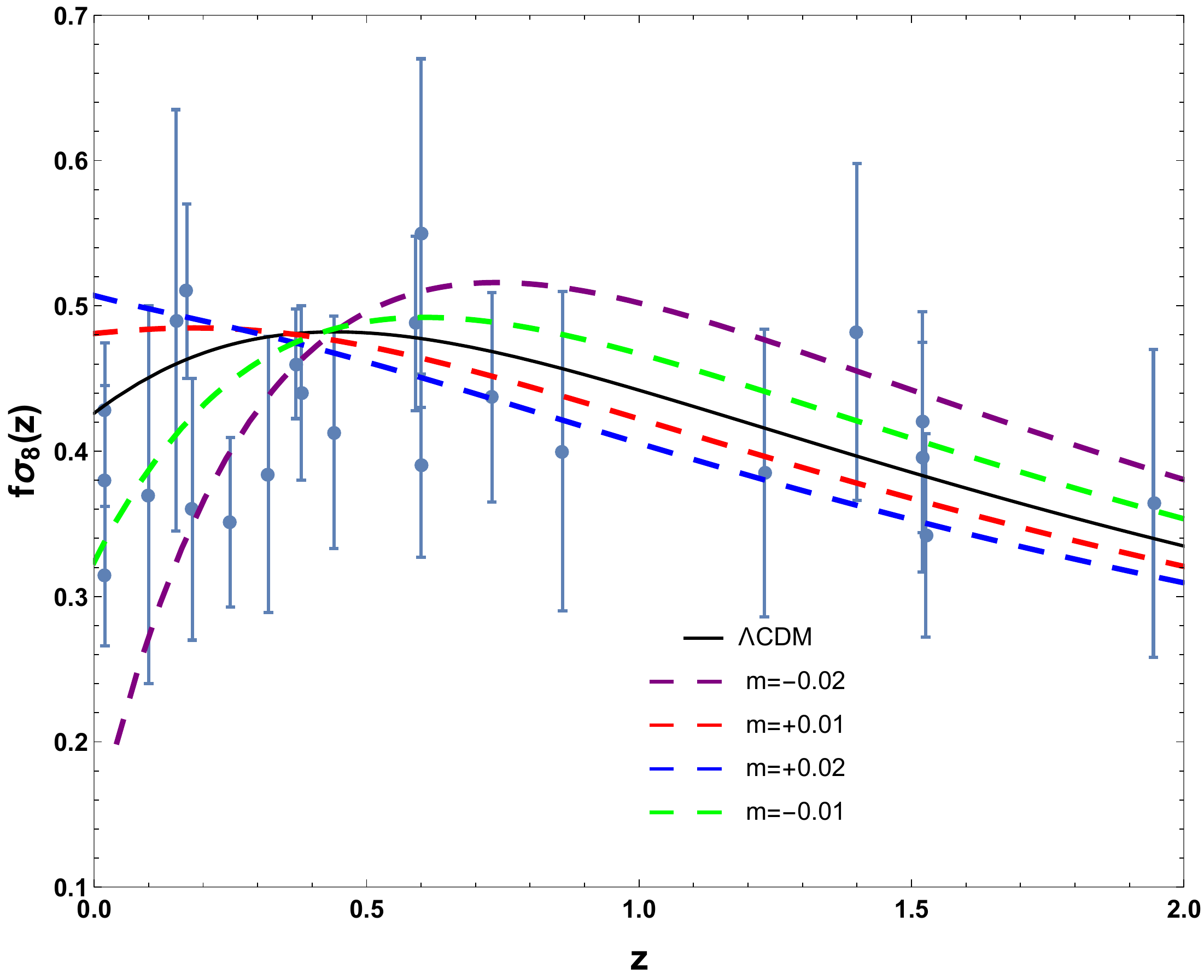}
\caption{Dependence of $f\sigma_8 (z)$ on $z$ for different values of the parameter $m$ with $\alpha = \beta = \mu= \nu = c_{\varepsilon}^{2} =0$.}
\label{fs8-m-0831}	
\end{figure}

\begin{figure}[H]
\centering
\includegraphics[width=0.6\textwidth]{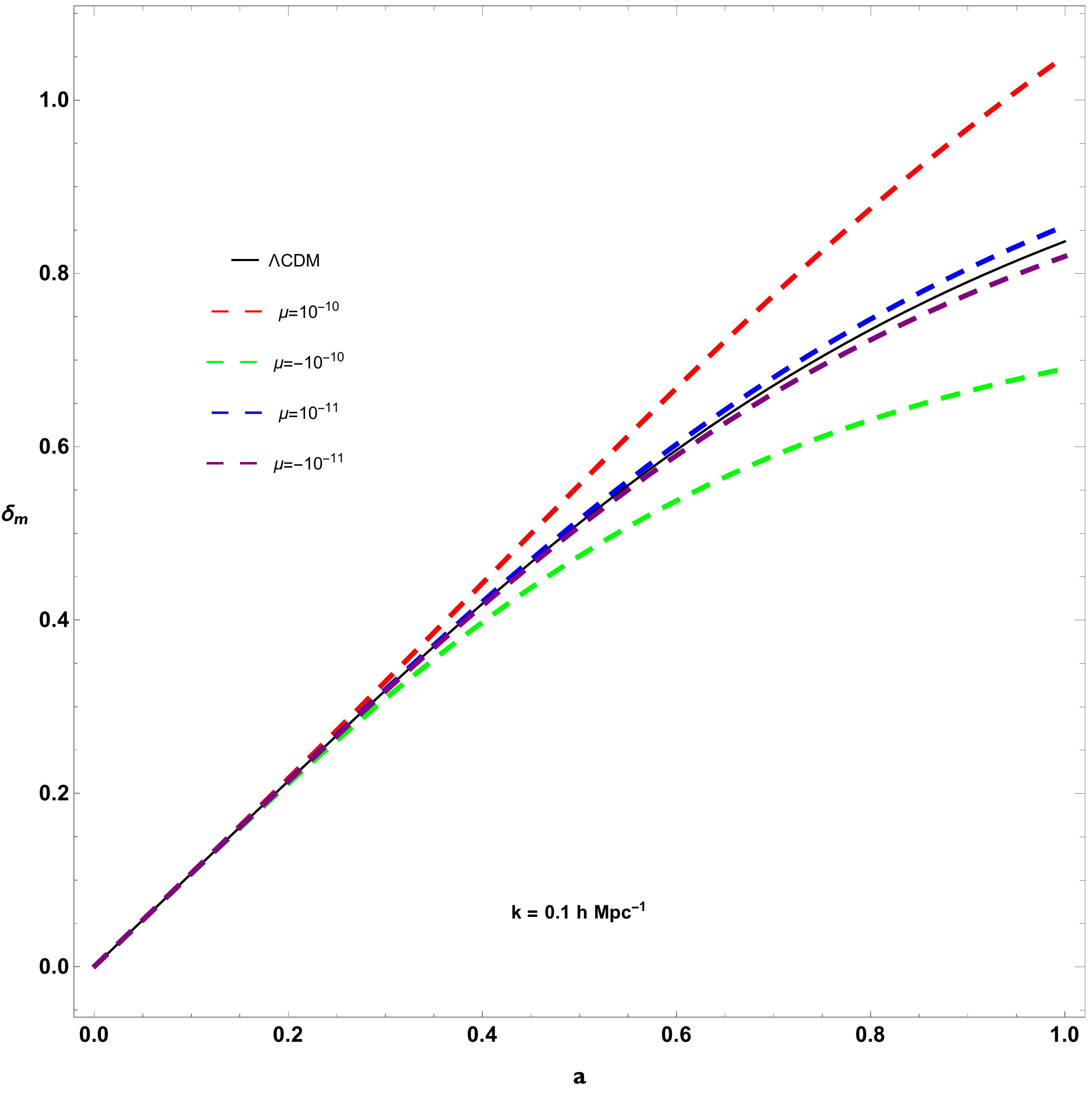}
\caption{Matter growth in the presence of anisotropic stresses ($m= \alpha = \beta = \nu = c_{\varepsilon}^{2} =0$).}
\label{delta-mu}	
\end{figure}

\begin{figure}[H]
\centering
\includegraphics[width=0.6\textwidth]{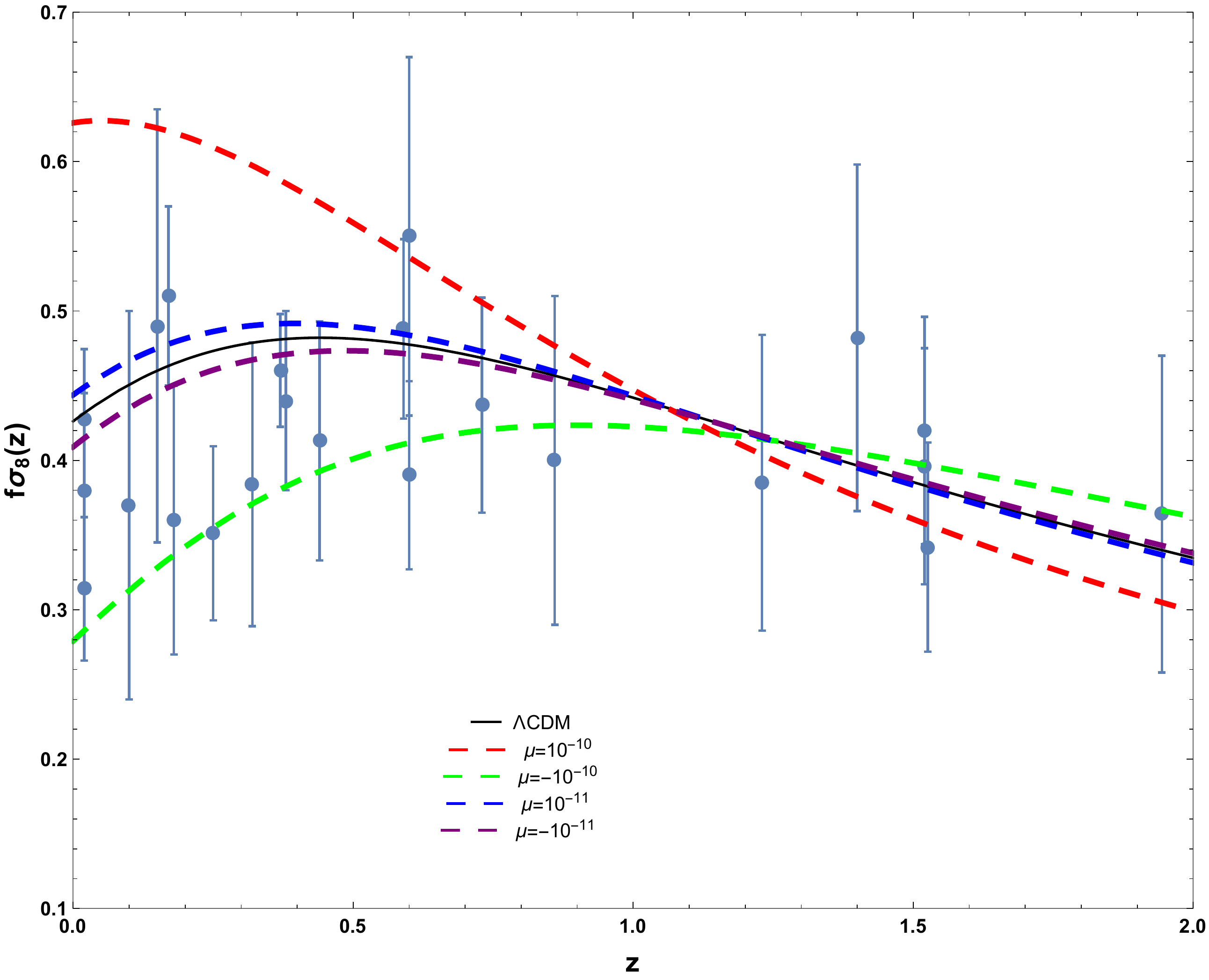}
\caption{Dependence of $f\sigma_8 (z)$ on $z$ in the presence of anisotropic stresses ($m= \alpha = \beta = \nu = c_{\varepsilon}^{2} =0$).}
\label{fs8-mu}
\end{figure}

\begin{figure}[H]
\centering
\includegraphics[width=0.6\textwidth]{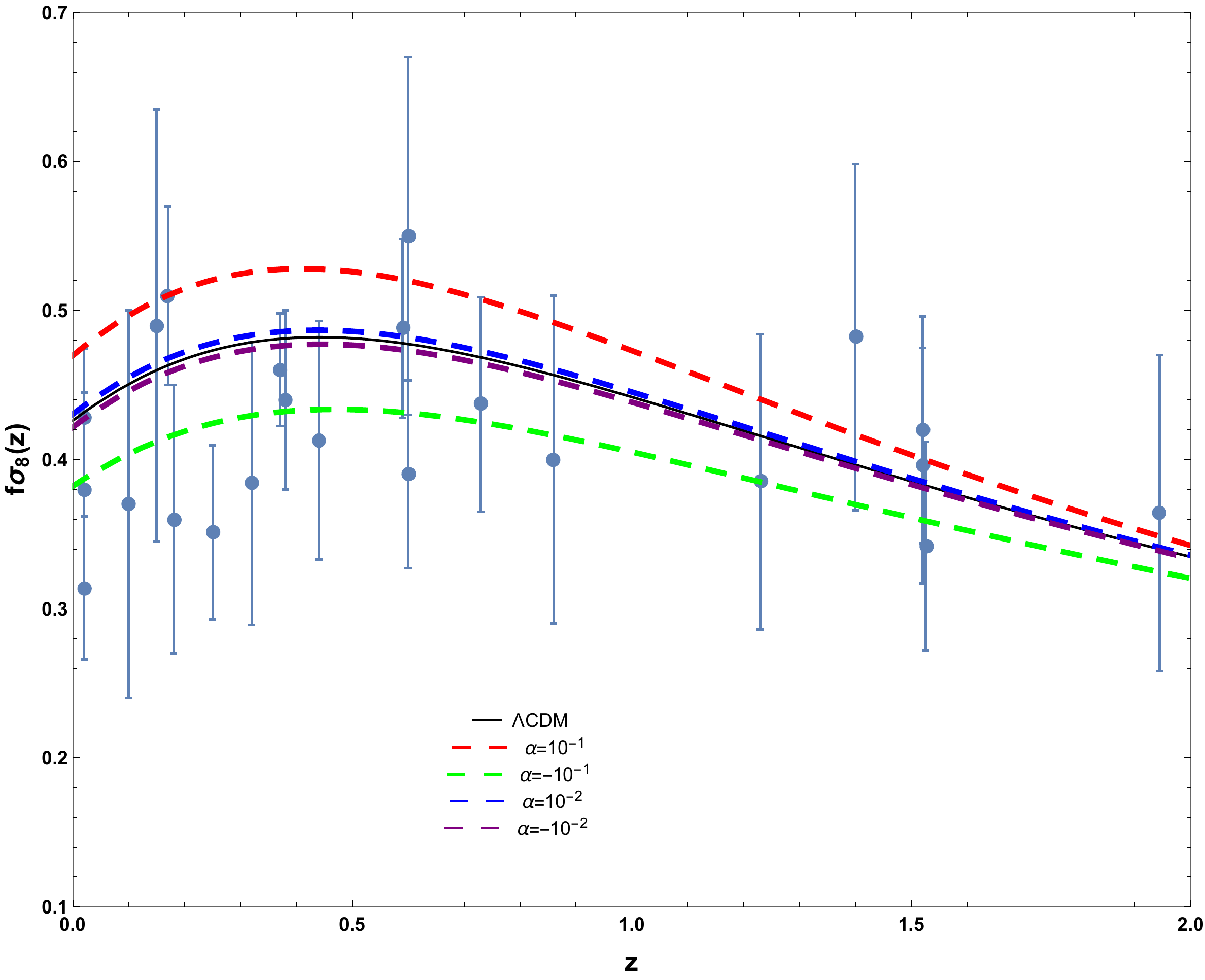}
\caption{Dependence of $f\sigma_8 (z)$ on $z$ in the presence of heat fluxes ($m = \beta = \mu  =\nu = c_{\varepsilon}^{2} =0$).}
\label{fs8-alpha}	
\end{figure}

Figure~\ref{delta-m} shows the fractional matter energy density perturbations $\delta_{m}$ as a function of the scale factor if only the parameter $m$ is  modified (different values of $m$ for  $\alpha = \beta = \mu= \nu =0$) with respect to the $\Lambda$CDM model ($m=0$).
 The dependence of the growth rate $f\sigma_8$ on the redshift for these values together with the data points is shown
in Figure~\ref{fs8-m-0831}.
Positive values of $m$  enhance the matter growth, negative values diminish it.
Figure~\ref{delta-mu}  visualizes the scale-factor dependence of the fractional matter density perturbations for different values of the anisotropic stress parameter $\mu$ for $m=\alpha = \beta = \nu =0$.
The corresponding growth rate $f\sigma_8$  for different values of $\mu$ is shown in Figure~\ref{fs8-mu}.
Deviations from the $\Lambda$CDM model require values of $\mu$ of the order of $|10^{-10}|$.
Finally, we demonstrate the impact of a non-vanishing heat flow, represented by the parameter $\alpha$, on the growth rate (see Figure~\ref{fs8-alpha}).
The heat flux basically causes a shift of the $\Lambda$CDM curve (upward for $\alpha >0$, downward for $\alpha<0$).

\section{Conclusions}
\label{discussion}

We have established a general phenomenological scheme for implementing (effective) non-equilibrium effects in a fluid description of the cosmological  dark sector.
This comprises both ``true" dissipative effects within Einstein's GR, assuming that the cosmic substratum behaves less simply than taken for granted in the usually applied  perfect-fluid approach, and those features which originate from a re-interpretation of geometric terms in modified gravitational theories in terms of effective fluid quantities.
Using a combination of
the fluid conservation equations with the Raychaudhuri equation for the expansion scalar, the complete  first-order, scalar perturbation dynamics have been reduced to a manifestly gauge-invariant coupled system of two second-order differential equations for the total and the relative energy--density perturbations.
We clarified how a  heat flux (effective or ``true") modifies the Poisson-type equation for the gravitational potential.
A characteristic feature of our approach consists in the introduction of phenomenological parameters with the purpose to make the fluid dynamical equations a closed system.
The relevant relations by which these coefficients are introduced are inspired by the manner that the speed of sound is conventionally introduced on a phenomenological basis.
In a standard perfect-fluid description of the Universe, such relation between the perturbations of (isotropic) pressure and energy density is required to close the system of perturbation equations. Here, we are generalizing this procedure by adding relations of a similar type which take into account anisotropic pressure and heat flux.
As in the case of a phenomenologically introduced sound speed, a derivation from an underlying fundamental theory is left open. Even for the sound speed parameter, an analytic microscopic justification does exist only in special cases.
This obvious shortcoming of the phenomenological theory is the price to pay for obtaining a robust, transparent and in large part analytical description of the inhomogeneous dynamics.

Our analysis is preliminary since it so far gives only a very rough account of the relevance of different (effective) dissipative phenomena on the growth of matter inhomogeneities during the cosmic history. At this point, also in view of the large error bars of the data, only order-of-magnitude estimates are possible.
Our approach allows for deviations from the standard model as long as these deviations are small.
Additional information is also needed to decide whether deviations from the standard model which are likely to be tiny, can be attributed to deviations from Einstein's GR or to a ``real" non-equilibrium nature of the cosmic substratum within GR.

In a general scalar-tensor theory, all the effective fluid quantities energy density, isotropic pressure, anisotropic pressure and heat flux are given in terms of the perturbations
of the scalar field \cite{madsen,faraoni18}. However, this dependence, while exact, is rather involved and, to obtain observationally relevant quantities, it needs numerical  implementation at a much earlier stage compared with the scheme presented here.
Of course, a final justification of this scheme  will require a sound microscopic foundation, a problem we hope to deal with for specific cases in future work.



\
\vspace{6pt}





{\bf Authors Contribution:} The authors contributed as a team to this work. Observations, numerical and statistical analysis: HESV and WCA, original draft: WZ.\\

{\bf Funding:}This research was funded by Conselho Nacional de Desenvolvimento Cient\'{\i}fico e Tecnol\'{o}gico (CNPq), FUNDA\c{C}\~{A}O ESTADUAL DE AMPARO \`{A} PESQUISA DO ESTADO DO ESP\'{I}RITO SANTO (FAPES) and
COORDENA\c{C}\~{A}O DE APERFEI\c{C}OAMENTO DE PESSOAL DE N\'{I}VEL SUPERIOR (CAPES).\\

{\bf Conflict of Interest:} The authors declare no conflict of interest. The founding sponsors had no role in the writing of the manuscript, and in the decision to publish the results.









\end{document}